\documentclass[12pt, onecolumn, final]{IEEEtran}
\usepackage{algorithm, algorithmic}
\usepackage{epsfig, graphics, color}
\usepackage{amsmath, amssymb, amsfonts, amsbsy, mathrsfs, bm}
\usepackage{cases}
\usepackage{enumerate, fancyvrb}
\usepackage{theorem}
\usepackage[compress]{cite}
\DeclareMathAlphabet{\mathsfsl}{OT1}{cmss}{m}{sl}

\ifCLASSOPTIONcompsoc
 \usepackage[tight, normalsize, sf, SF]{subfigure}
\else
 \usepackage[tight, footnotesize]{subfigure}
\fi

\usepackage{array}
\newcolumntype{L}[1]{>{\raggedright\let\newline\\\arraybackslash\hspace{0pt}}m{#1}}
\newcolumntype{C}[1]{>{\centering\let\newline\\\arraybackslash\hspace{0pt}}m{#1}}
\newcolumntype{R}[1]{>{\raggedleft\let\newline\\\arraybackslash\hspace{0pt}}m{#1}}

\graphicspath{{./figure/}}

\title{Distributed Local Linear Parameter Estimation using Gaussian SPAWN}
\author{Mei Leng, Wee Peng Tay, Tony~Q.S.~Quek, and Hyundong Shin
\thanks{M. Leng and W.P. Tay are with the Nanyang Technological University, Singapore. (e-mail:$\{$lengmei,wptay$\}$@ntu.edu.sg).}
\thanks{T.Q.S. Quek is with the Information Systems Technology and Design Pillar, SUTD, Singapore. (e-mail:qsquek@ieee.org).}
\thanks{H. Shin is with the Department of Electronics and Radio Engineering, Kyung Hee University, Yongin-si, Gyeonggi-do, 446-701, Korea (e-mail: hshin@khu.ac.kr).}}

\newtheorem{theorem}{Theorem}
\newtheorem{proposition}{Proposition}
\newtheorem{lemma}{Lemma}
\newtheorem{corollary}{Corollary}
\newtheorem{assumption}{Assumption}

\newcommand{\bB}{\mathbf{B}}

\newcommand{\bD}{\mathbf{D}}
\newcommand{\bE}{\mathbf{E}}

\newcommand{\bH}{\mathbf{H}}

\newcommand{\bO}{\mathbf{O}}

\newcommand{\bQ}{\mathbf{Q}}

\newcommand{\bW}{\mathbf{W}}

\newcommand{\rA}{\mathrm{A}}
\newcommand{\rB}{\mathrm{B}}
\newcommand{\rC}{\mathrm{C}}
\newcommand{\rD}{\mathrm{D}}
\newcommand{\rE}{\mathrm{E}}
\newcommand{\rF}{\mathrm{F}}
\newcommand{\rG}{\mathrm{G}}
\newcommand{\rH}{\mathrm{H}}
\newcommand{\rJ}{\mathrm{J}}
\newcommand{\rK}{\mathrm{K}}

\newcommand{\rP}{\mathrm{P}}
\newcommand{\rQ}{\mathrm{Q}}
\newcommand{\rR}{\mathrm{R}}

\newcommand{\rT}{\mathrm{T}}
\newcommand{\rI}{\mathrm{I}}
\newcommand{\rU}{\mathrm{U}}
\newcommand{\rV}{\mathrm{V}}
\newcommand{\rW}{\mathrm{W}}
\newcommand{\rX}{\mathrm{X}}
\newcommand{\rY}{\mathrm{Y}}

\newcommand{\brH}{{\mathrm{H}}}

\newcommand{\cF}{\mathcal{F}}
\newcommand{\cR}{\mathcal{R}}

\newcommand{\bd}{\mathbf{d}}

\newcommand{\bs}{\mathbf{s}}

\newcommand{\stB}{\mathcal{B}}
\newcommand{\stE}{\mathcal{E}}
\newcommand{\stG}{\mathcal{G}}
\newcommand{\stS}{\mathcal{S}}

\newcommand{\dd}{\mathrm{d}}

\newcommand{\bpsi}{\bm{\psi}}

\newcommand{\bnu}{\bm{\nu}}
\newcommand{\bmu}{\bm{\mu}}
\newcommand{\tmu}{\tilde{\bmu}}
\newcommand{\bomega}{\bm{\omega}}

\newcommand{\E}{\mathbb{E}}
\newcommand{\N}[2]{{\mathcal{N}\left(#1\ ;\ #2\right)}}
\newcommand{\ofrac}[1]{{\frac{1}{#1}}}
\newcommand{\tc}[1]{^{(#1)}}
\newcommand{\abs}[1]{{\left| #1 \right|}}
\newcommand{\norm}[1]{{\left\lVert #1 \right\rVert}}

\newcommand{\EE}[1]{{\mathbb{E}\left\{ #1 \right\}}}
\newcommand{\p}[1]{{p\left(#1\right)}}
\newcommand{\condp}[2]{{p\left(#1 \left\lvert#2\right. \right)}}

\newcommand{\vect}[1]{\mathrm{vec}\left\{#1\right\}}

\newcommand{\approptoinn}[2]{\mathrel{\vcenter{
  \offinterlineskip\halign{\hfil$##$\cr
    #1\propto\cr\noalign{\kern2pt}#1\sim\cr\noalign{\kern-2pt}}}}}

\newcommand{\appropto}{\mathpalette\approptoinn\relax}

\begin{document}
\maketitle

\begin{abstract}
We consider the problem of estimating local sensor parameters, where the local parameters and sensor observations are related through linear stochastic models. Sensors exchange messages and cooperate with each other to estimate their own local parameters iteratively. We study the Gaussian Sum-Product Algorithm over a Wireless Network (gSPAWN) procedure, which is based on belief propagation, but uses fixed size broadcast messages at each sensor instead. Compared with the popular diffusion strategies for performing network parameter estimation, whose communication cost at each sensor increases with increasing network density, the gSPAWN algorithm allows sensors to broadcast a message whose size does not depend on the network size or density, making it more suitable for applications in wireless sensor networks. We show that the gSPAWN algorithm converges in mean and has mean-square stability under some technical sufficient conditions, and we describe an application of the gSPAWN algorithm to a network localization problem in non-line-of-sight environments. Numerical results suggest that gSPAWN converges much faster in general than the diffusion method, and has lower communication costs, with comparable root mean square errors.
\end{abstract}


\section{Introduction}
A wireless sensor network (WSN) consists of many sensors or nodes capable of on-board sensing, computing, and communications. WSNs are used in numerous applications like environmental monitoring, pollution detection, control of industrial machines and home appliances, event detection, and object tracking \cite{Akyildiz2007,Bulusu2005,Tay2009,Tay2008}. In distributed estimation \cite{Hong2007, Zhu2010, Speranzon2006}, sensors cooperate with each other by passing information between neighboring nodes, which removes the necessity of transmitting local data to a central fusion center. Distributed estimation schemes hence have the advantages of being scalable, robust to node failures, and are more energy efficient due to the shorter sensor-to-sensor communication ranges, compared with centralized schemes which require transmission to a fusion center. It also improves local estimation accuracy \cite{Gholami2012, Dardari2008}. For example, by cooperating with each other to perform localization, nodes that have information from an inadequate number of anchors can still successfully self-localize \cite{Wymeersch2008,Wymeersch2009}. 

In this paper, we consider distributed local linear parameter estimation in a WSN in which each sensor cooperates with its neighbors to estimate its own local parameter, which is related to its own observations as well as its neighbors' observations via a linear stochastic model. This is a special case of the more general problem in which sensors cooperate to estimate a \emph{global} vector parameter of interest, as the local parameters can be collected into a single vector parameter. Many distributed estimation algorithms for this problem have been investigated in the literature, including consensus strategies \cite{Boyd2006,Aysal2009,Zhu2011,Kriegleder2013}, the incremental least mean square (LMS) algorithm \cite{Lopes2007, Cattivelli2011}, the distributed LMS consensus-based algorithm \cite{Schizas2009} and diffusion LMS strategies (see the excellent tutorial \cite{Sayed2013a} and the references therein). The consensus strategies are relatively less computationally intensive for each sensor. However, their performance in terms of the convergence rate and mean-square deviation (MSD) are often not as good as the diffusion LMS strategies \cite{Tu2012}. The incremental LMS algorithm requires that sensors set up a Hamiltonian path through the network, which is impractical for a large WSN. 

To ensure mean and mean-square convergence, the diffusion LMS strategy requires that a pre-defined step size in each diffusion update is smaller than twice the reciprocal of the maximum eigenvalue of the system matrix covariance \cite{Sayed2013,Sayed2013a}. Since this eigenvalue is not known a priori, a very small step size is thus typically chosen in the diffusion LMS strategy. However, this leads to slow convergence rates, resulting in higher communication costs. Furthermore, the diffusion strategy is not specifically designed to perform local sensor parameter estimation in a WSN. For example, when sensors need to estimate their own clock skews and offsets in network synchronization \cite{Leng2011}, the diffusion LMS strategy either requires that every node estimates the same global parameter, which is a collection of all the sensor local clock skews and offsets, or at least transmits estimates of the clock skews and offsets of all its neighbors (cf.\ Section \ref{subsect:compare_ATC} for a detailed discussion). In both cases, communication cost for each sensor per iteration increases with the density of the network, and does not make full use of the broadcast nature of wireless communications in a WSN. The distributed LMS consensus-based algorithm of \cite{Schizas2009} requires the selection of a set of bridge sensors, and the passing of several messages between neighboring sensors, which again does not utilize the broadcast nature of a WSN. It is also more computationally complex than the diffusion algorithm but with better MSD performance.

We therefore ask if there exists a distributed estimation algorithm with similar MSD performance as the diffusion algorithms, and that allows sensors to \emph{broadcast a fixed size message}, regardless of the network size or density, to all its neighbors? Our work suggests that the answer is affirmative for a somewhat more restrictive data model than that used in the LMS literature \cite{Sayed2013a}, and can be found in the Sum-Product Algorithm over a Wireless Network (SPAWN) method, first proposed by \cite{Wymeersch2009}. The SPAWN algorithm is based on belief propagation \cite{Kschischang2001, Bishop2006}; the main differing characteristic is that the \emph{same} message is broadcast to all neighboring nodes by each sensor, in contrast to traditional belief propagation where a different message is transmitted to each neighbor. The reference \cite{Wymeersch2009} however does not address the issue of error convergence in the SPAWN algorithm, and it is well known that loopy belief propagation may not converge \cite{Pearl1988,Johnson2006, Weiss2001}.

In this paper, we consider an adaptive version of the Gaussian SPAWN (gSPAWN) algorithm for distributed local linear parameter estimation. We assume that sensors have only local observations with respect to its neighbors, such as the pairwise distance measurement, the relative temperature with respect to one another, and the relative clock offset between two nodes. Similar to \cite{Cattivelli2010, Takahashi2010, Sayed2013}, we assume that the observations follow a linear model. Our main contribution in this paper is the derivation of sufficient conditions for the mean and MSD convergence of the gSPAWN algorithm. Note that although the gSPAWN algorithm is based on Gaussian belief propagation, the methods of \cite{Weiss2001, Johnson2006} for analyzing the convergence of Gaussian belief propagation in a loopy graph do not apply due to the difference in messages transmitted by each node in gSPAWN versus that in Gaussian belief propagation. In fact, due to the broadcast nature of the messages in gSPAWN, our analysis is simpler than that in \cite{Weiss2001, Johnson2006}.

As an example, we apply the gSPAWN algorithm to cooperative self-localization in non-line-of-sight (NLOS) multipath environments. To the best of our knowledge, most distributed localization methods \cite{Srirangarajan2008, Ihler2005, Zhu2011} consider only line-of-sight (LOS) signals, because NLOS environments introduce non-linearities in the system models, and measurement noises can no longer be modeled as Gaussian random variables. In this application, we assume that individual propagation paths can be resolved, and we adopt a ray tracing model to characterize the relationship between sensor locations, range and angle measurements \cite{Miao2007, Seow2008, Xie2009}. When all scatterers in the environment are either parallel or orthogonal to each other, we show that the sensor location estimates given by the gSPAWN algorithm converges in the mean to the true sensor positions. We compare the performance of our algorithm with that of a peer-to-peer localization method and the diffusion LMS strategy, with numerical results suggesting that the gSPAWN algorithm has better average accuracy and convergence rate.

The rest of this paper is organized as follows. In Section \ref{Section:System}, we define the system model. In Section \ref{Section:Algo}, we describe the gSPAWN algorithm and compare it to the diffusion algorithm. We provide sufficient conditions for mean convergence and mean-square stability of the gSPAWN algorithm, and numerical comparison results in Section \ref{Section:convergence}. We then show an application of the gSPAWN algorithm to network localization in NLOS environments in Section \ref{Section:Case}. Finally, we summarize and conclude in Section \ref{Section:Conclude}.

\emph{Notations:} We use upper-case letters to represent matrices and lower-case letters for vectors and scalars. Bold faced symbols are used to denote random variables. The conjugate of a matrix $A$ is $\bar{A}$. The transpose and conjugate transpose of $A$ are denoted as $A^T$ and $A^*$, respectively. The minimum and maximum non-zero singular values of $A$ are denoted as $\underline\delta(A)$ and $\bar\delta(A)$, respectively. The maximum absolute eigenvalue or spectral radius of $A$ is denoted as $r(A)$. When $A$ is Hermitian, we have $r(A) = \bar\delta(A)$. We write $A \succeq B$ and $A \succ B$  if $A - B$ is positive semi-definite and positive definite, respectively. If all entries of $B-A$ are non-negative, we write $A \leq B$. The operation  $A \otimes B$ is the Kronecker product between $A$ and $B$. The vector $\vect{A}$ is formed by stacking the columns of $A$ together into a column vector. The matrix $\mathrm{diag}(A_1,\ldots,A_n)$ is a block diagonal matrix consisting of the sub-matrices $A_1, \ldots, A_n$ on the main diagonal. The symbol $\rI_m$ represents a $m\times m$ identity matrix. We use $0_{m}$ to denote a $m\times 1$ vector of all zeroes.
The operator $\E$ denotes mathematical expectation. The density function of a multivariate Gaussian distribution with mean $\mu$ and covariance $\rP$ is given by $\N{\cdot}{\mu, \rP}$.

\section{System model}\label{Section:System}
In this section, we describe our system model, and present some assumptions that we make throughout this paper. We consider a network of $n+1$ sensors $\{0, \ldots, n\}$, where each sensor $i$ wants to estimate a $d\times 1$ parameter $s_i$. Sensor $0$ is set to be the reference node in the network, and $s_0$ is a global reference level for the other sensor parameters. For example, in a localization problem, we are often interested to find the relative locations of nodes in the network with respect to (w.r.t.) a reference node. In the sensor network synchronization problem, we want to estimate the relative clock offset and skew of each sensor w.r.t. to a reference sensor. In this paper, we use the terms ``sensors'' and ``nodes'' interchangeably. We say that sensors $i$ and $j$ are neighbors if they are able to communicate with each other. A network consisting of all sensors therefore corresponds to a graph $\stG=(\stE, \stS)$, where the set of vertices $\stS=\{0,\ldots,n\}$, and the set of edges $\stE$ consists of communication links in the network. We let the set of neighbors of sensor $i$ be $\stB_i$ (which may include sensor $i$ itself or not, depending on the application data model). 

Sensors interact with neighbors and estimate their local parameters through in-network processing. In most applications, each sensor can obtain only local measurements w.r.t.\ its neighbors via communication between each other. Examples include the pseudo distance measurement between two sensors when performing wireless ranging, and the pairwise clock offset when performing clock synchronization. For each sensor $i$, and neighbor $j \in \stB_i$, we consider the data model given by
\begin{align}
\bd_{ij}\tc{l} = \rG_{ij}s_i - \bH_{ij}\tc{l}s_j + \bomega_{ij}\tc{l}, \label{linear}
\end{align}
where $\bd_{ij}\tc{l}$ is the $m\times 1$ vector measurement obtained at sensor $i$ w.r.t.\ sensor $j$ at time $l$, $\rG_{ij}$ is a known $m \times d$ system coefficient matrix, $\bH_{ij}\tc{l}$ is an observed $m\times d$ regression matrix at time $l$, and $\bomega_{ij}\tc{l}$ is a measurement noise of dimensions $m\times 1$. We note that this data model is a special case of the widely studied LMS data model (see e.g., \cite{Sayed2013a}), since we can describe \eqref{linear} by using a global parameter $s = [s_0^T, \ldots, s_n^T]^T$ with appropriate stacking of the measurements $\bd_{ij}\tc{l}$. On the other hand, if every sensor $i$ is interested in the same global parameter $s_i = s$, we have from \eqref{linear} that $\bd_{ij}\tc{l} = \mathbf{U}_{ij}\tc{l} s + \bomega_{ij}\tc{l}$, where $\mathbf{U}_{ij}\tc{l} = \rG_{ij} - \bH_{ij}\tc{l}$ is the regression matrix in \cite{Sayed2013a}.\footnote{The data model in \cite{Sayed2013a} assumes that $m=1$, but can be easily extended to $m > 1$.} Although we have assumed for convenience that all quantities have the same dimensions across sensors, our work can be easily generalized to the case where measurements and parameters have different dimensions at each sensor. We have the following assumptions similar to \cite{Sayed2013a} on our data model.

\begin{assumption}\label{assumpt:model1}\
\begin{enumerate}[(i)]
	\item The regression matrices $\bH_{ij}\tc{l}$ are independent over sensor indices $i$ and $j$, and over time $l$. They are also independent of the measurement noises $\bomega_{ij}\tc{l}$ for all $i,j$, and $l$. For all $i,j$, $\bH_{ij}\tc{l}$ is stationary over $l$.
	\item The measurement noises $\bomega_{ij}\tc{l}$ are stationary with zero mean, and $\E[\bomega_{ij}\tc{l}(\bomega_{ij}\tc{l})^*] = \rC_{ij}$, where $\rC_{ij}$ is a positive semi-definite matrix.
	\item\label{it:GG} For every $i = 1,\ldots,n$, the matrix $\sum_{j\in\stB_i} \rG_{ij}^*\rG_{ij}$ is positive definite.
\end{enumerate}
\end{assumption}

In the LMS framework, each sensor $i$ seeks to estimate $s_i$ so that the overall network mean square error,
\begin{align*}
\sum_{i=1}^n\sum_{j\in\stB_i} \EE{\norm{\bd_{ij}\tc{1} - \rG_{ij}s_i + \bH_{ij}\tc{1}s_j}^2}
\end{align*}
is minimized. Our goal is to design a distributed algorithm to perform local parameter estimation, which makes full use of the broadcast nature of the wireless medium over which sensors communicate. In particular, the messages broadcast by each sensor do not depend on the number of neighbors of the sensor or the network size. This is achieved by the gSPAWN algorithm. However, in order to ensure convergence of the gSPAWN algorithm, we need the following additional technical assumptions as well as Assumption \ref{assumpt:model3}, which will be discussed later in Section \ref{subsect:Belief_means_converge}. This makes our system model more restrictive than that considered in \cite{Sayed2013a}. 

\begin{assumption}\label{assumpt:model2}\
\begin{enumerate}[(i)]
	\item\label{it:G} The network graph $\mathcal{G}$ is connected, and $\mathcal{G}\backslash\{0\}$ consists of strongly connected components.\footnote{The notation $\mathcal{G}\backslash\{0\}$ denotes a graph with node $0$ and its incident edges removed.} 
	\item\label{it:s0} The reference sensor $0$ has an a priori estimate $\hat{s}_0 = s_0 + w_0$ of its parameter $s_0$, where $w_0$ is a zero mean Gaussian random variable with covariance matrix $\alpha_0 \rI_d$.
	\item\label{it:GH_sv}  Let $\brH_{ij} = \E[\bH_{ij}\tc{l}]$ for all $l\geq 1$, $\eta_i^2 = \underline{\delta}(\sum_{j\in\stB_i}\rG_{ij}^*\rG_{ij})$, and $\rho_i^2 = \max_{j\in\stB_i} \bar{\delta}(\rH_{ij}\rH_{ij}^*)$. For all $i \geq 1$, we have $\rho_i^2 \leq \eta_i^2/\abs{\stB_i}$ if $\abs{\stB_i} \leq 2$, and $\rho_i^2 \leq \eta_i^2/3$ if $\abs{\stB_i} \geq 3$. 
\end{enumerate}
\end{assumption}

Assumption \ref{assumpt:model2}\eqref{it:G} does not result in much loss of generality. If the graph $\mathcal{G}$ is not connected, the same analysis can be repeated on each connected component of $\mathcal{G}$. Furthermore, in most practical applications, sensor measurements are symmetric, i.e., if sensor $i$ can obtain a measurement w.r.t.\ sensor $j$, then sensor $j$ can also obtain a measurement w.r.t.\ sensor $i$. Assumption \ref{assumpt:model2}\eqref{it:s0} also holds in our applications of interest, where typically, $\hat{s}_0 = s_0$ is taken to be a known reference value with $\alpha_0 = 0$. For example, when localizing sensors w.r.t.\ to the reference node 0, there is no loss in generality in assuming that the reference node is at $\hat{s}_0 = s_0 = 0_d$, the origin of the frame of reference. Assumption \ref{assumpt:model2}\eqref{it:GH_sv} on the other hand restricts our data model to those in which the system coefficient matrices $\rG_{ij}$ and expected regression matrices $\brH_{ij}$ do not differ by too much.

\section{The Gaussain SPAWN Algorithm}\label{Section:Algo}

In this section, we briefly review the SPAWN algorithm, and derive the adaptive gSPAWN algorithm. Since belief propagation is a Bayesian estimation approach, we will \emph{temporarily} view the parameters $s_1, \ldots, s_n$ as random variables with uniform priors in this section. Suppose that we make the observations $\bO\tc{l} = (\bd_{ij}\tc{l}, \bH_{ij}\tc{l})_{i\geq1,j\in\stB_i}$ at a time $l$. In the Bayesian framework, we are interested to find values $s_1,\ldots,s_n$ that maximize the \emph{a posteriori} probability 
\begin{align*}
\p{\{s_i\}_{i=1}^n \mid \bO\tc{l}},
\end{align*}
where the notation $\p{\mathbf{x} \mid \mathbf{y}}$ denotes the conditional probability of $\mathbf{x}$ given $\mathbf{y}$.\footnote{Note the abuse of notation here: $\mathbf{x}$ is used to denote both the random variable and its realization.} Since the distributions of $\bd_{ij}\tc{l}$ and $\bd_{ji}\tc{l}$ depend only on $s_i$ and $s_j$, we have from \eqref{linear} that
\begin{align}
    \p{\{s_i\}_{i=1}^n \mid \bO\tc{l}}
    & \propto \prod_{(i,j) \in \stE} \condp{\bd_{ij}\tc{l}, \bd_{ji}\tc{l}}{s_i, s_j, \bH_{ij}\tc{l}, \bH_{ji}\tc{l}} \nonumber\\
    & = \prod_{(i,j) \in \stE} \condp{\bd_{ij}\tc{l}}{s_i,s_j, \bH_{ij}\tc{l}} \condp{\bd_{ji}\tc{l}}{s_j,s_i, \bH_{ji}\tc{l}}. \label{sp}
\end{align}

\begin{figure}[!t]
\centering
\includegraphics[width=0.4\textwidth]{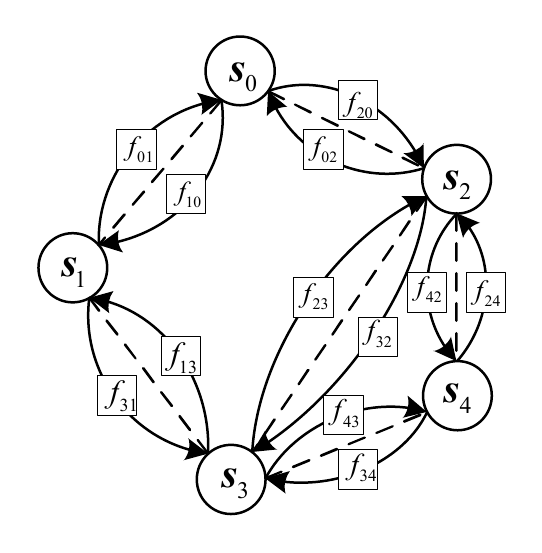}
\caption{A factor graph of a network with $5$ sensors, where a dashed line indicates that a communication path exists between the corresponding two sensors, and the arrows indicate the direction of message flows.}\label{Figure-FG}
\end{figure}

Let $f_{ij}(s_i,s_j)=\condp{\bd_{ij}\tc{l}}{s_i,s_j, \bH_{ij}\tc{l}}$ and $f_{ji}(s_i,s_j)=\condp{\bd_{ji}\tc{l}}{s_j,s_i, \bH_{ji}\tc{l}}$. We construct a factor graph using $\{s_i\}_{i=0}^n$ as the variable nodes, and for each pair of neighboring sensors $i$ and $j$ with $(i,j) \in \stE$, we connect their corresponding variable nodes using the two factor nodes $f_{ij}$ and $f_{ji}$. See Figure \ref{Figure-FG} for an example, where each random variable is represented by a circle and each factor node is represented by a square. We note that our factor graph construction is somewhat untypical compared to those used in traditional loopy belief propagation, where usually a single factor $f_{ij}(s_i,s_j)f_{ji}(s_i,s_j)$ is used between the two variable nodes $s_i$ and $s_j$. The reason we separate it into two factors is to allow us to design an algorithm that allows sensors to broadcast the same message to all neighboring nodes, as described below.

To apply the sum-product algorithm on a loopy factor graph, we need to define a message passing schedule \cite{Bishop2006}. We adopt a fully parallel schedule here. In each iteration, all factor nodes send messages to their neighboring variable nodes in parallel, followed by messages from variable nodes to neighboring factor nodes in parallel. We constrain the updates by allowing messages to flow in only one direction. Specifically, for every $i$ and for all $j \in \stB_i$, $s_i$ sends messages only to $f_{ji}$ and not to $f_{ij}$, and $f_{ij}$ sends messages only to $s_i$ and not to $s_j$. The two types of messages sent between variable and factor nodes at the $k$th iteration are the following:
\begin{itemize}
\item{Each factor node $f_{ij}$ passes a message $h\tc{k}_{f_{ij} \rightarrow s_i}(s_i)$ to the variable node $s_i$ representing $f_{ij}$'s belief of $s_i$'s state. This message is given by
\begin{align}
h\tc{k}_{f_{ij} \rightarrow s_i}(s_i) = \int \condp{\bd_{ij}\tc{l}}{s_i, s_j, \bH_{ij}\tc{l}} b^{(k-1)}_{j}(s_j) \ \dd s_j. \label{message}
\end{align}}
\item{Each variable node $s_i$ \emph{broadcasts} a belief to all neighboring factor nodes $\{f_{ji}\}_{j \in \stB_i}$, and its belief is given by 
\begin{align}
b_{i}\tc{k}(s_i) = \prod_{j \in \stB_i } h\tc{k}_{f_{ij} \rightarrow \bs_i}(\bs_i). \label{belief}
\end{align}
}
\end{itemize}
The above scheme is called the SPAWN algorithm by \cite{Wymeersch2009}. Note that the same belief message from the variable node $s_i$ is broadcast to all factor nodes $\{f_{ji}\}_{j \in \stB_i}$, since node $s_i$ receives only from factor node $f_{ij}$ and the belief $b\tc{k}_i(s_i)$ does not include any messages from $f_{ji}$ in the previous iterations. The communication cost therefore \emph{does not depend on the size or density of the network}, making this scheme more suitable for a WSN. In the following, we derive the exact messages passed when $\condp{\bd_{ij}\tc{l}}{s_i, s_j, \bH_{ij}\tc{l}}$, for all $i, j$, are assumed to be Gaussian distributions. In addition, we allow the observations to be updated at each iteration so that the algorithm becomes adaptive in nature. We call this the gSPAWN algorithm. Although Gaussian distributions are assumed in the derivation of the gSPAWN algorithm, we show in Section \ref{Section:convergence} that this assumption is not required for mean and MSD convergence of the algorithm.\footnote{This is analogous to the philosophy of using the B.L.U.E. for parameter estimation with non-Gaussian system models.}

\subsection{Derivation of Beliefs and Messages for gSPAWN}\label{Subsection:Derivation}

We first consider the message from the reference node $0$ to a neighboring node $i$. The reference node's belief is held constant over all iterations. From Assumption \ref{assumpt:model2}\eqref{it:s0}, we have $b_0^{(l)}(s_0) = \N{s_0}{\bmu_0\tc{l}, \rP_0\tc{l}}$, where $\bmu_0\tc{l} = \hat{s}_0$ and $\rP_0\tc{l} = \alpha_0\rI_d$, for all $l\geq 0$. The message from the factor $f_{i0}$ to sensor $i$ is given by
\begin{align}
h_{f_{i0} \rightarrow \bs_i}^{(l)}(\bs_i) & = \int \condp{\bd_{i0}\tc{l}}{s_i, s_0, \bH_{i0}\tc{l}} b_0^{(l)}(s_0) \ \dd s_0 \nonumber \\
& \appropto \N{\rG_{i0} s_i}{\bnu_{i0}\tc{l},\Pi_{i0}\tc{l}}. \label{S0Si}
\end{align}
where $\bnu_{i0}\tc{l} = \bd_{i0}\tc{l} + \bH_{i0}\tc{l}\bmu_0\tc{l}$, $\Pi_{i0}\tc{l} = \rC_{i0} + \brH_{i0}\rP_0\tc{l}\brH_{i0}^*$, and the symbol $\appropto$ means approximately proportional to. The approximation in \eqref{S0Si} arises because we have replaced $\bH_{i0}\tc{l}$ with $\brH_{i0}$ in $\Pi_{i0}\tc{l}$.

For all nodes $i$ other than the reference node $0$, we set the initial belief $b_i^{(0)}(s_i)$ of node $i$ to be a Gaussian distribution with mean $\mu_j\tc{0} = 0_d$ and covariance
\begin{align}\label{initial_P}
\rP_i\tc{0} & = \left\{
\begin{array}{ll}
4\alpha \rI_d, & \textrm{if  $|\mathcal{B}_{i}| =1$}, \\
3\alpha \rI_d, & \textrm{if  $|\mathcal{B}_{i}| = 2$},\\
5\alpha/3 \rI_d, & \text{otherwise,}
\end{array}
\right.
\end{align}
where $\alpha$ is a positive number chosen to be sufficiently large so that $\alpha \geq \alpha_0$, and $\alpha \geq \frac{\max_{i,j} r(\rC_{ij})}{\min_{i}\rho_{i}^2}$, where $\rho_i^2 = \max_{j\in\stB_i} \bar{\delta}(\rH_{ij}\rH_{ij}^*)$ (cf.\ Assumption \ref{assumpt:model2}\eqref{it:GH_sv}). A larger initial variance is used for those nodes with less neighbors because such nodes have relatively less information than other nodes in the network. We will see later in the proof of Theorem \ref{theorem:P} that the chosen multipliers for the cases where there are not more than two neighbors are the right ones.

Consider the $l$th iteration of the gSPAWN algorithm. After $l-1$ iterations, the belief of node $j$  is $b_j^{(l-1)}(s_j) = \N{s_j}{\bmu_j^{(l-1)}, \rP_j^{(l-1)}}$. The message $b_j^{(l-1)}(s_j)$ is a Gaussian distribution, so only the mean $\bmu_j^{(l-1)}$ and covariance matrix $\rP_j^{(l-1)}$ need to be passed to node $i$, which then computes
\begin{align*}
h_{f_{ij} \rightarrow s_i}^{(l)}(s_i) & = \int \condp{\bd_{ij}\tc{l}}{s_i,s_j, \bH_{ij}\tc{l}} b_j^{(l-1)}(s_j) \ \dd \bs_j \\
& \appropto \N{\rG_{ij}s_i}{\bnu_{ij}\tc{l-1}, \Pi_{ij}\tc{l-1}},
\end{align*}
where
\begin{align}
\bm{\nu}_{ij}^{(l-1)} & = \bd_{ij}\tc{l}+\bH_{ij}\tc{l} \bmu_{j}^{(l-1)}, \label{invmu} \\
\Pi_{ij}\tc{l-1} & = \rC_{ij} +\brH_{ij}\rP\tc{l-1}_{j}\brH_{ij}^*. \label{inv}
\end{align}

We obtain the belief of node $i$ at the $l$th iteration from (\ref{belief}). Since all the messages in the product in \eqref{belief} are Gaussian distributions, we have $b_i^{(l)}(s_i) = \N{s_i}{\bmu_i^{(l)}, \rP_i^{(l)}}$ with
\begin{align}
\rP_i^{(l)} & = \left(\sum_{j \in \stB_i} \rG_{ij}^*\left(\Pi_{ij}\tc{l-1}\right)^{-1}\rG_{ij}\right)^{-1} \label{P}
\end{align}
and
\begin{align}
\bm{\mu}_i^{(l)} = \rP_i^{(l)} \sum_{j \in \stB_i} \rG_{ij}^*\left(\Pi_{ij}\tc{l-1}\right)^{-1} \bm{\nu}_{ij}^{(l-1)}. \label{mu}
\end{align}
We will show in Lemma \ref{lemma:P-positive} that $\rP_j\tc{l}$ is positive definite for all $j=1,\ldots,n$ and $l\geq0$, so \eqref{P} is well defined.

\begin{algorithm}[!bt]
\caption{gSPAWN: Gaussian Sum-Product Algorithm over a Wireless Network} \label{algorithm}
\begin{algorithmic}[1]
\STATE{\textbf{Initialization}:}
\STATE{For the reference node $0$, set $\bmu_0\tc{l}=\hat{s}_0$ and $\rP_0\tc{l} = \alpha_0 \rI_d$, for all $l \geq 0$.}
\STATE{For $i=1,\ldots,n$, let $\bmu_i^{(0)} = 0_d$ and choose $\rP_i\tc{0}$ according to \eqref{initial_P}.}
\FOR{\textrm{the $l^{\textrm{th}}$ iteration}}
\STATE{\textbf{nodes} $i = 1 : n$} \textbf{in parallel}
\STATE{broadcast the current belief $b^{(l-1)}_i(s_i)$ to neighboring nodes;}
\STATE{receive $b^{(l-1)}_j(s_j)$ from neighboring nodes $j$, where $j \in \stB_i$;}
\STATE{update its belief as $b^{(l)}_i(\bs_i) \sim \N{s_i}{\bmu^{(l)}_i,\rP^{(l)}_i}$ with
\begin{align*}
\rP_i^{(l)} = \left(\sum_{j \in \stB_i} \rG_{ij}^*\left(\Pi_{ij}\tc{l-1}\right)^{-1}\rG_{ij}\right)^{-1},
\end{align*}
and
\begin{align*}
\bmu_i^{(l)} = \rP_i^{(l)} \sum_{j \in \stB_i} \rG_{ij}^*\left(\Pi_{ij}\tc{l-1}\right)^{-1} \bnu_{ij}^{(l-1)},
\end{align*}
where $\bnu_{ij}^{(l-1)}$ and $\Pi_{ij}\tc{l-1}$ are given in \eqref{invmu} and \eqref{inv} respectively.}
\STATE{\textbf{end parallel}}
\ENDFOR
\end{algorithmic}
\end{algorithm}

The gSPAWN algorithm is formally presented in Algorithm \ref{algorithm}. Note that factor nodes are virtual, and are introduced only to facilitate the derivation of the algorithm. In practice, the update is performed at each individual node directly. At the end of $l$ iterations, node $i$ estimates its parameter $s_i$ by maximizing the belief $b^{(l)}_i(s_i)$ with respect to $s_i$. Since $b^{(l)}_i(s_i)$ is a Gaussian distribution, the local estimator for $s_i$ is given by $\hat{s}^{(l)}_i = \bmu\tc{l}_i$, and has a covariance of $\rP_i\tc{l}$. 

\subsection{Comparison with Diffusion Adapt-Then-Combine}\label{subsect:compare_ATC}

The gSPAWN algorithm is reminiscent of the Adapt-Then-Combine (ATC) diffusion algorithm, first proposed by \cite{Cattivelli2010}. We present the ATC algorithm in our context of local parameter estimation,\footnote{We consider only the version of ATC in which there is no exchange of sensor observations with neighbors. We also do not discuss the Combine-Then-Adapt diffusion algorithm as this is known to perform worse than ATC in general \cite{Cattivelli2010}.} and discuss the differences with the gSPAWN algorithm. As the ATC algorithm assumes that the data model at every sensor is based on a common global parameter, we stack the local sensor parameters into a single global parameter $s = [s_0^T, \ldots, s_n^T]^T$, with the data model at sensor $i$ given by
\begin{align}
\bd_i\tc{l} = \bW_i\tc{l} s + (\bomega_i\tc{l})^T, \label{dLMS}
\end{align}
where $\bomega_i\tc{l} = \left[(\bomega_{i1}\tc{l})^T, \ldots, (\bomega_{in}\tc{l})^T\right]$, and $\bW_i\tc{l}$ is a $(n+1)\times (n+1)$ block matrix consisting of $d\times d$ blocks, with the $(j,k)$-th block for $0\leq j, k \leq n$ being
\begin{align*}
\bW_i\tc{l}(j,k) =
\begin{cases}
\rG_{ij},  & \textrm{if $k=i$ and $j\in\stB_i\backslash\{i\}$},\\
\rG_{ii} - \bH_{ii}\tc{l}, & \textrm{if $k=i$ and $j = i \in\stB_i$},\\
-\bH_{ij}\tc{l}, & \textrm{if $k=j$ and $j \in \stB_i\backslash\{i\}$},\\
0\cdot \rI_d, & \textrm{otherwise}. 
\end{cases} 
\end{align*}
The ATC update at time $l$ at sensor $i$ consists of the following steps:
\begin{align}
\bpsi_i\tc{l} &= \bs_i\tc{l-1} + \xi_i\cdot (\bW_i\tc{l})^*(\bd_i\tc{l} - \bW_i\tc{l}\bs_i\tc{l-1}), \label{ATC_psi}\\
\bs_i\tc{l} &= \sum_{j\in\stB_i} a_{ij} \bpsi_{j}\tc{l}, \label{ATC_s}
\end{align}
where $\bpsi_i\tc{l}$ and $\bs_i\tc{l}$ are the intermediate and local estimates of $s$ at sensor $i$ respectively, $\xi_i$ is a step size chosen to be sufficiently small, and $(a_{ij})$ are non-negative weights with $\sum_j a_{ij} = 1$ for all $i$. The following remarks summarize the major differences of gSPAWN and ATC.
\begin{enumerate}[(i)]
	\item In the $l$-th iteration of the ATC update \eqref{ATC_psi}-\eqref{ATC_s}, each sensor $i$ first computes $\bpsi_i\tc{l}$ and broadcasts it so that its neighbors can perform the combination step \eqref{ATC_s}. The message $\bpsi_i\tc{l}$ is an intermediate estimate of the global parameter $s$, and is of size $dn \times 1$. Each sensor $i$ needs to know the total number of nodes $n$ in the network, which determines the size of its broadcast message. Its communication cost therefore increases linearly with $n$. In contrast, in the gSPAWN algorithm, sensor $i$ broadcasts a fixed size message $(\bmu_i\tc{l}, \rP_i\tc{l})$, which does not depend on the size of the network.
	\item Alternatively, instead of broadcasting the message $\bpsi_i\tc{l}$ in full, sensor $i$ can choose to broadcast only those components of $\bpsi_i\tc{l}$ that it updates, which correspond to the parameters $\{s_j : j \in \stB_i\}$. In this case, the sensor has to add identity headers in its broadcast message so that the receiving sensors know which parameters each component corresponds to. In a static network, this can be done once but in a mobile network, the header information has to be re-broadcast each time the network topology changes. The message itself also has size linear in the number of neighbors of the sensor. Therefore, the communication cost per sensor per iteration increases with the density of the network. We note that the communication cost for a network using gSPAWN does not depend on the network node density.
	\item In the gSPAWN algorithm, the matrix weights in \eqref{mu} need not be non-negative, unlike those in \eqref{ATC_s}. Moreover, the weights in \eqref{mu} evolve over the iterations, and reflect each sensor's confidence in its belief. The weights for the ATC algorithm we have presented are fixed, although adaptive combination weights for ATC have also been proposed in \cite{Sayed2013a} but no convergence analysis is available. 
	\item In \eqref{ATC_psi}, to ensure mean and mean-square convergence, the step size $\xi_i$ is chosen so that
	\begin{align}
	0 < \xi_i < \frac{2}{r(\rR_{i})}, \label{ATCstep}
	\end{align}
where $\rR_i=\E[\bW_i^*\bW_i]$ \cite{Sayed2013}. Since $r(\rR_i)$ is in general not known a priori, the step size $\xi_i$ is chosen to be very small in most applications. The convergence rate of the ATC algorithm is controlled by the step size, with a smaller step size leading to a lower convergence rate. In the gSPAWN algorithm, the parameter $\alpha$ in \eqref{initial_P} plays an analogous role, but it, in effect, does not control the convergence rate since it appears in both $\Pi_{ij}\tc{l-1}$ and $\rP_i\tc{l}$ in \eqref{mu}, and can be chosen as large as desired (see Section \ref{Subsection:Derivation}).
	\item In Section \ref{Section:convergence}, we show the mean convergence and mean-square stability of gSPAWN under some technical assumptions, which make our model less general than the LMS model used in the ATC algorithm. The advantages of gSPAWN over ATC as described above are only valid under these assumptions. Nevertheless, these assumptions hold for several important sensor network applications including distributed sensor localization, which we discuss in Section \ref{Section:Case}, and distributed sensor clock synchronization \cite{Leng2011}. We also remark that gSPAWN is more suitable for \emph{local} parameter estimation, while ATC has lower communication costs in the case of \emph{global} network parameter estimation if the number of parameters is fixed, since gSPAWN needs to broadcast covariance information as well as parameter estimates.
\end{enumerate}

\section{Convergence analysis of \normalfont{gSPAWN}} \label{Section:convergence}

In this section, we show that the covariance matrices $\rP_i^{(l)}$ in the gSPAWN algorithm converge, and establish sufficient conditions for the belief means to converge to the true parameter values. We also show that the MSD converges under additional assumptions. Although the gSPAWN algorithm follows the general framework of belief propagation, its message passing is different from traditional belief propagation. Moreover, we consider an adaptive version of the algorithm in this paper. Therefore, the general convergence analysis in \cite{Pearl1988,Johnson2006, Weiss2001} cannot be applied here. 

\subsection{Convergence of Covariance Matrices}
We first state several elementary results. The first lemma is from \cite{Sarhan2010}, and shows that a non-increasing sequence of positive definite matrices (in the order defined by $\succeq$) converges. The proof of Lemma \ref{lemma-matrix-inv} can be found in \cite{Milliken1977}.

\begin{lemma}\label{lemma-matrix-seq}
If the sequence $\{\rX^{(l)}\}_{l = 1}^{+\infty}$ of positive definite matrices is non-increasing, i.e., $\rX^{(l)} \succeq \rX^{(l+1)}$ for $l = 1, 2, \cdots$, then this sequence converges to a positive semi-definite matrix.
\end{lemma}

\begin{lemma} \label{lemma-matrix-inv}
If the matrices $\rX$ and $\rY$ are positive definite, then $\rX \succeq \rY$ if and only if $\rY^{-1} \succeq \rX^{-1}$.
\end{lemma}

\begin{lemma}\label{lemma:P-positive}
Suppose that Assumption \ref{assumpt:model1}\eqref{it:GG} holds. For all $i=1,\ldots,N$, and for all $l \geq 0$, $\rP_i\tc{l}$ is positive definite.
\end{lemma}
\begin{IEEEproof}
We prove the claim by induction on $l$. From \eqref{initial_P}, the claim holds for $l=0$. Suppose the claim holds for $l-1$, we have that $\Pi_{ij}\tc{l-1}$ is positive definite for all $j \in \mathcal{B}_i$. Therefore, the inverse $(\Pi_{ij}\tc{l-1})^{-1}$ is also positive definite, and from \eqref{P} and Assumption \ref{assumpt:model1}\eqref{it:GG}, since $\sum_{j\in\stB_i}\rG_{ij}^*\rG_{ij}$ has full rank, we obtain that $\rP_i\tc{l}$ is positive definite. The induction is now complete, and the lemma is proved.
\end{IEEEproof}

The following result shows that the covariance matrices of the beliefs at each variable node in the factor graph converges.
\begin{theorem}\label{theorem:P}
Suppose that Assumptions \ref{assumpt:model1} and \ref{assumpt:model2} hold. Then, the covariance matrices $\rP_i^{(l)}$ at every node $i$ in the gSPAWN algorithm converges, i.e., there exists unique positive semi-definite matrices $\{\rP_i\tc{\infty}\}_{i = 1}^{n}$ such that for all $i=1,\ldots,n$,
\begin{align}
\lim_{l \rightarrow \infty} \rP^{(l)}_i = \rP_i\tc{\infty}. \label{Pconv}
\end{align}
\end{theorem}
\begin{IEEEproof}
All nodes are updated using \eqref{P} and \eqref{mu}. Since the covariance matrix $\rC_{ij}$ is positive definite, we can perform eigen decomposition on $\rC_{ij}$. Let $\rC_{ij} = \rV_{ij}^* \rD_{ij} \rV_{ij}$, where $\rV_{ij}$ is a unitary matrix, and $\rD_{ij}$ is a diagonal matrix with positive entries. Let $\rK_i^{(l)} = \left(\rP_i^{(l)}\right)^{-1}$. From \eqref{inv} and \eqref{P}, it can be shown that
\begin{align} \label{K}
\rK_i\tc{l} = \sum_{j\in\stB_i} \rG_{ij}^*\rV_{ij}^*\left(\rD_{ij} + \rV_{ij}\brH_{ij}\rP_j\tc{l-1}\brH_{ij}^*\rV_{ij}^*\right)^{-1} \rV_{ij}\rG_{ij}.
\end{align}
Further performing singular value decomposition on $\rG_{ij}$ and $\brH_{ij}$, we have $\rG_{ij}=\rU_{ij}\rA_{ij}\tilde{\rU}_{ij}^*$ and $\brH_{ij} = \rW_{ij}\rB_{ij} \tilde{\rW}_{ij}^*$, where $\rU_{ij}$, $\tilde{\rU}_{ij}$, $\rW_{ij}$, and $\tilde{\rW}_{ij}$ are unitary matrices, and $\rA_{ij}$ and $\rB_{ij}$ are diagonal matrices with corresponding singular values on the diagonal. We have,
\begin{align}
\rK_i\tc{1} & = \sum_{j\in\stB_i} \tilde{\rU}_{ij}\rA_{ij}^*\rU_{ij}^*\rV_{ij}^*\left(\rD_{ij} + \rV_{ij}\rW_{ij}\rB_{ij}\rP_j\tc{l-1}\rB_{ij}^*\rW_{ij}^*\rV_{ij}^*\right)^{-1}\rV_{ij}\rU_{ij}\rA_{ij}\tilde{\rU}_{ij}^*. \label{Kil}
\end{align}
We show by induction on $l$ that $\rK_i\tc{l}$ is non-decreasing for all $i=0,\ldots,N$. Since the reference node $0$ does not update its belief, this clearly holds for $i=0$. In the following, we prove the claim for $i=1,\ldots,n$. Since $\alpha \geq \max_{i,j}r(\rD_{ij})/\rho_i^2$, it follows that $\rD_{ij} \preceq \alpha\rho_i^2\rI_m$ for all $i,j$. From Assumption \ref{assumpt:model2}\eqref{it:GH_sv}, we also have $\bar\delta(\rH_{ij}\rH_{ij}^*) \leq \rho_i^2$, and hence $\rB_{ij}\rB_{ij}^* \preceq \rho_i^2 \rI_m$ for all $i,j$. Using these two properties, we now consider different cases for the size of $\stB_i$.

Case $|\stB_i| = 1$: From Assumption \ref{assumpt:model2}\eqref{it:GH_sv}, we have $\rho_i^2 \leq \eta_i^2$. We consider the cases where the reference node $0$ is a neighbor of node $i$ or not separately. Suppose that the node $0$ is the neighbor of node $i$. Then, since $\rP_0\tc{0}=\alpha_0\rI_2$, we have from \eqref{Kil} that
\begin{align*}
\rK_i\tc{1} & \succeq \frac{\eta_i^2}{(\alpha + \alpha_0)\rho_i^2} \rI_d \\
& \succeq  \frac{\eta_i^2}{2\alpha\rho_i^2} \rI_d \\
& \succeq \ofrac{4\alpha} \rI_d = \rK_i\tc{0},
\end{align*}
where the second inequality follows from $\alpha_0 \leq \alpha$. If node $0$ is not a neighbor of node $i$, suppose that $\stB_i = \{j\}$, where $j \ne 0$. From Assumption \ref{assumpt:model2}\eqref{it:G}, we have $j \ne i$, and $|\stB_j| \geq 2$, yielding
\begin{align*}
\rK_i\tc{1} & \succeq \frac{\eta_i^2}{\alpha (\rho_i^2 + 3\rho_i^2)} \rI_d \\
& \succeq \frac{1}{4\alpha} \rI_d = \rK_i\tc{0}.
\end{align*}

Case $|\stB_i| = 2$: From Assumption \ref{assumpt:model2}\eqref{it:GH_sv}, we have $\rho_i^2 \leq \eta_i^2/2$. Denote $\stB_i = \{j_1, j_2\}$, we have $|\stB_{j_1}| =1$ and $|\stB_{j_2}|=2$ in the worst case, and it follows from \eqref{Kil} that
\begin{align*}
\rK_i\tc{1} & \succeq \ofrac{\alpha}\frac{\rG_{ij_1}^*\rG_{ij_1}}{\rho_i^2 + 4\rho_i^2}
+ \ofrac{\alpha}\frac{\rG_{ij_2}^*\rG_{ij_2}}{\rho_i^2 + 3 \rho_i^2} \\
& = \ofrac{\alpha} \frac{4\rG_{ij_1}^*\rG_{ij_1} + 5\rG_{ij_2}^*\rG_{ij_2}}{20 \rho_i^2}\\
& \succeq \ofrac{\alpha} \frac{\eta_i^2}{5\rho_i^2}  \rI_d\\
& \succeq \frac{1}{3\alpha}\rI_d  = \rK_i\tc{0}.
\end{align*}

Case $\abs{\stB_i} \geq 3$: From \eqref{Kil}, we have
\begin{align*}
\rK_i\tc{1} & \succeq \sum_{j\in\stB_i} \tilde{\rU}_{ij}\rA_{ij}^*\rU_{ij}^*\rV_{ij}^*\left(\rD_{ij} + 4\alpha\rV_{ij}\rW_{ij}\rB_{ij}\rB_{ij}^*\rW_{ij}^*\rV_{ij}^*\right)^{-1}\rV_{ij}\rU_{ij}\rA_{ij}\tilde{\rU}_{ij}^*.
\end{align*}
From Assumption \ref{assumpt:model2}\eqref{it:GH_sv}, we have $\rho_i^2 \leq \eta_i^2/3$, and we obtain
\begin{align*}
\rK_i\tc{1} & \succeq \frac{\sum_{j\in\stB_i} \tilde{\rU}_{ij}\rA_{ij}^*\rA_{ij}\tilde{\rU}_{ij}^*}{5 \alpha \rho_i^2}\\
& \succeq \frac{1}{\alpha} \frac{\eta_i^2}{5\rho_i^2} \rI_d \\
& \succeq \frac{3}{5\alpha} \rI_d = \rK_i\tc{0}.
\end{align*}
Therefore, we have shown that $\rK_i\tc{1} \succeq \rK_i\tc{0}$ for all nodes $i\geq 1$.

Suppose that $\rK_j\tc{l} \succeq \rK_j\tc{l-1}$ for all $j$. From Lemmas  \ref{lemma-matrix-inv} and \ref{lemma:P-positive}, we have $\rP_{j}\tc{l-1} \succeq \rP_{j}\tc{l}$, and it follows that $\rV_{ij}\brH_{ij}\rP_j\tc{l-1}\brH_{ij}^*\rV_{ij}^* \succeq \rV_{ij}\brH_{ij}\rP_j\tc{l}\brH_{ij}^*\rV_{ij}^*$. Applying Lemma \ref{lemma-matrix-inv}, we have
\begin{align*}
\left(\rD_{ij} + \rV_{ij}\brH_{ij}\rP_j\tc{l}\brH_{ij}^*\rV_{ij}^*\right)^{-1} \succeq \left(\rD_{ij} + \rV_{ij}\brH_{ij}\rP_j\tc{l-1}\brH_{ij}^*\rV_{ij}^*\right)^{-1},
\end{align*}
which together with \eqref{K} yields
\begin{align*}
\rK_i\tc{l+1}\succeq \rK_i\tc{l}.
\end{align*}
This completes the induction, and the claim that $\rK_i\tc{l}$ is non-decreasing in $l$ for each $i$ is now proven. This implies that $\rP_i\tc{l}$ is non-increasing in $l$. The theorem now follows from Lemma \ref{lemma-matrix-seq}, and the proof is complete.
\end{IEEEproof}

Theorem \ref{theorem:P} shows that the covariance matrices at every node converges to the positive semi-definite matrices $\{\rP_i\tc{\infty}\}_{i = 1}^{n}$, which is the fixed point solution to the set of equations
\begin{align}\label{Pinfty} 
\left(\rP_i\tc{\infty}\right)^{-1} = \sum_{j\in\stB_i} \rG_{ij}^*\left(\rC_{ij} + \brH_{ij}\rP_j\tc{\infty}\brH_{ij}^*\right)^{-1} \rG_{ij},
\end{align}
for all $i \geq 1$, with $\rP_0\tc{\infty} = \alpha_0 \rI_d$.

\subsection{Convergence of Belief Means}\label{subsect:Belief_means_converge}

In this subsection, we provide a technical assumption that together with Assumptions \ref{assumpt:model1} and \ref{assumpt:model2} ensure that the gSPAWN algorithm converges in mean. We then provide sufficient conditions that satisfy the given technical assumption. Using the covariance matrices $\{\rP_i\tc{\infty}\}_{i = 0}^{n}$, which can be derived using \eqref{Pinfty}, we let $\rQ\tc{\infty}$ be a $dn \times dn$ matrix consisting of $n \times n$ blocks of $d \times d$ submatrices, where for $1 \leq i, j \leq n$, the $(i,j)$-th block is
\begin{align}
\rQ\tc{\infty}(i,j) =
\begin{cases}
\rP\tc{\infty}_i \rG_{ij}^*\left(\rC_{ij} + \brH_{ij}\rP_j\tc{\infty}\brH_{ij}^*\right)^{-1}\brH_{ij}, & \text{if } j \in \stB_i, \\
0\cdot \rI_d, & \textrm{otherwise}.
\end{cases} \label{Qinfty}
\end{align}

\begin{assumption}\label{assumpt:model3}
The spectral radius $r(\rQ\tc{\infty})$ is less than 1.
\end{assumption}
Assumption \ref{assumpt:model3}, together with Assumptions \ref{assumpt:model1} and \ref{assumpt:model2}, are sufficient for the mean convergence (Theorem \ref{theorem:mu}) of the gSPAWN algorithm. If the observations $(\bd_{ij}\tc{l}, \bH_{ij}\tc{l})_{i,j,l}$ are scalars, then it can be shown that Assumption \ref{assumpt:model3} holds under quite general conditions, see \cite{Leng2011}. However, for vector observations, $\rQ\tc{\infty}$ in general does not have spectral radius less than 1 even when $\brH_{ij} = \rG_{ij}$ for all $i,j$ (a numerical example is provided in Appendix \ref{appendix:counterexample}). To verify Assumption \ref{assumpt:model3} requires a priori knowledge of the sensor network architecture, which may not be practical. To overcome this difficulty, we present local sufficient conditions in the following proposition that ensure Assumption \ref{assumpt:model3} is satisfied.

\begin{proposition}\label{prop:sufficient}
Suppose that Assumptions \ref{assumpt:model1} and \ref{assumpt:model2} hold. In addition, there exists a unitary matrix $\rU$ so that for all $i, j = 0, \ldots, n$, there exists a unitary matrix $\rV_{ij}$ such that 
\begin{enumerate}[(i)]
	\item $\rG_{ij} = \rV_{ij} \rA_{ij} \rU^*$, where $\rA_{ij}$ is a diagonal matrix with non-negative real numbers on the diagonal;
	\item $\brH_{ij} = \rV_{ij} \rB_{ij} \rU^*$, where $\rB_{ij}$ is a diagonal matrix with non-negative real numbers on the diagonal, with $\rB_{ij} \leq \rA_{ij}$; and
	\item $\rC_{ij} = \rV_{ij} \rD_{ij} \rV_{ij}^*$, where $\rD_{ij}$ is a diagonal matrix with positive real numbers on the diagonal.
\end{enumerate}
Then, Assumption \ref{assumpt:model3} holds.
\end{proposition}
\begin{IEEEproof}
We first show, by induction on $l$, that for all $i \geq 0$ and $l \geq 0$, $\rP_i\tc{l}$ has the form $\rU \Lambda_{i}\tc{l} \rU^*$, where $\Lambda_i\tc{l}$ is a non-negative diagonal matrix. This is trivially true for all $i \geq 0$ when $l = 0$. It is also true for $i=0$ and for all $l \geq 1$ since the reference node $0$ does not update its belief. Suppose that the claim holds for $l-1$. Then we have for $i \geq 1$,
\begin{align}
\left(\rP_i\tc{l}\right)^{-1} &= \sum_{j \in \stB_i} \rG_{ij}^*\left(\rC_{ij} + \brH_{ij}\rP_j\tc{l-1}\brH_{ij}^*\right)^{-1}\rG_{ij} \nonumber\\
& = \rU \underbrace{\sum_{j \in \stB_i} \rA_{ij} \left(\rD_{ij} + \rB_{ij}\Lambda_j\tc{l-1}\rB_{ij} \right)^{-1} \rA_{ij}}_{\triangleq \Lambda_i\tc{l}} \rU^*, \label{Pl}
\end{align}
so the claim holds since $\rA_{ij} \left(\rD_{ij} + \rB_{ij}\Lambda_j\tc{l-1}\rB_{ij} \right)^{-1} \rA_{ij}$ is a diagonal matrix for all $i, j$. This implies that for all $i\geq 0$, $\rP_i\tc{\infty} = \rU \Lambda_{i}\tc{\infty} \rU^*$, for some non-negative diagonal matrix $\Lambda_i\tc{\infty}$ as $l\rightarrow \infty$. We hence have from \eqref{Qinfty} for $j\in\stB_i$,
\begin{align}
\rQ\tc{\infty}(i,j) & = \rP\tc{\infty}_i \rG_{ij}^*\left(\rC_{ij} + \brH_{ij}\rP_j\tc{\infty}\brH_{ij}^*\right)^{-1}\brH_{ij} \nonumber\\
& =  \rU \left(\Lambda_{i}\tc{\infty}\right)^{-1} \rA_{ij}  \left(\rD_{ij} + \rB_{ij}\Lambda_j\tc{\infty}\rB_{ij} \right)^{-1} \rB_{ij} \rU^*. \label{eq:Qij}
\end{align}
Let $\tilde{\rQ}$ be a a $dn \times dn$ matrix consisting of $n \times n$ blocks of $d \times d$ submatrices, where the $(i,j)$-th block is $(\Lambda_{i}\tc{\infty})^{-1}\rA_{ij} (\rD_{ij} + \rB_{ij}\Lambda_j\tc{\infty}\rB_{ij} )^{-1} \rB_{ij}$ for $1 \leq i, j \leq n$ and $j\in\stB_i$. Then from \eqref{eq:Qij}, $r(\rQ\tc{\infty})=r(\tilde{\rQ})$. We have
\begin{align}
\sum_{\substack{j \in \stB_i \\ j \ne 0}} \tilde{\rQ}(i,j) \leq \left(\Lambda_{i}\tc{\infty}\right)^{-1} \sum_{\substack{j \in \stB_i \\ j \ne 0}} \rA_{ij}  \left(\rD_{ij} + \rB_{ij}\Lambda_j\tc{\infty}\rB_{ij} \right)^{-1} \rA_{ij}. \label{eq:tQij}
\end{align}
From \eqref{Pl}, the right hand side (R.H.S.) of \eqref{eq:tQij} is equal to $\rI_d$ if $0 \notin \stB_i$, and $\prec \rI_d$ if $0 \in \stB_i$. This implies that $\tilde{\rQ}$ is a non-negative sub-stochastic matrix. From Assumption \ref{assumpt:model2}, $\tilde{\rQ}$ can be expressed, by relabeling the nodes if necessary, as a block diagonal matrix where each diagonal block is irreducible (when the reference node $0$ is removed from the network, the network breaks into irreducible components represented by each diagonal block). Applying the Perron-Frobenius theorem \cite{Seneta1981} on each block, we obtain the proposition.
\end{IEEEproof}

Let $\tmu_i\tc{l} = \bmu_i\tc{l} - s_i$. Substituting \eqref{invmu} and \eqref{inv} into \eqref{mu}, we obtain
\begin{align*}
\tmu_i\tc{l} & = \rP_i\tc{l} \sum_{j \in \stB_i} \rG_{ij}^*\left(\Pi_{ij}\tc{l-1}\right)^{-1} \bH_{ij}\tc{l} \tmu_j\tc{l-1} + \rP_i\tc{l} \sum_{j \in \stB_i} \rG_{ij}^*\left(\Pi_{ij}\tc{l-1}\right)^{-1} \bomega_{ij}\tc{l}. 
\end{align*}
Denote $\tmu\tc{l} = \left[\tmu_1^{(l)}, \ldots, \tmu_n^{(l)}\right]^T$, and $\bm{\Omega}\tc{l} = \left[\bomega_1\tc{l}, \cdots, \bomega_n\tc{l}\right]^{T}$, where $\bomega_i\tc{l} = \left[(\bomega_{i1}\tc{l})^T, \ldots, (\bomega_{in}\tc{l})^T\right]$, we have
\begin{align}\label{mu_recursive}
\tmu\tc{l} = \bQ\tc{l}\tmu\tc{l-1} +\rR\tc{l}\bm{\Omega}\tc{l},
\end{align}
where $\bQ\tc{l}$ is a $dn \times dn$ matrix consisting of $n \times n$ blocks of $d \times d$ submatrices, with the $(i,j)$-th block for $1\leq i,j \leq n$ being
\begin{align}
\bQ\tc{l}(i,j) =
\begin{cases}
\rP_i\tc{l} \rG_{ij}^*\left(\Pi_{ij}\tc{l-1}\right)^{-1} \bH_{ij}\tc{l}, & \text{if } j \in \stB_i, \\
0\cdot \rI_d, & \textrm{otherwise}.
\end{cases} \label{Q}
\end{align}
The matrix $\rR\tc{l} = \rE (\rI_n \otimes \tilde{\rQ}\tc{l})$, where $\tilde{\rQ}\tc{l}$ is the same as $\bQ\tc{l}$ except that it is missing the $ \bH_{ij}\tc{l}$ terms on the R.H.S.\ of \eqref{Q}, and $\rE$ is a selection matrix defined as
\begin{align*}
\rE =
\begin{bmatrix}
e_{1}^T \otimes \rI_d \\
\vdots \\
{e}_{(i-1)n + i}^T \otimes \rI_d \\
\vdots \\
{e}_{n^2}^T \otimes \rI_d
\end{bmatrix},
\end{align*}
with $e_i$ being a $n^2 \times 1$ vector where there is a 1 at the $i$-th component, and all other entries are zero. From Theorem \ref{theorem:P}, the following lemma follows easily.

\begin{lemma} \label{lemma:Q} Let $\rQ\tc{l} = \EE{\bQ\tc{l}}$. Suppose that Assumptions \ref{assumpt:model1}-\ref{assumpt:model3} hold. Then, $\lim_{l \to \infty} \rQ^{(l)} = \rQ\tc{\infty}$ in \eqref{Qinfty}, and for all $l$ sufficiently large, we have $r(\rQ\tc{l}) < 1$. Furthermore, there exists a finite constant $c$ such that for all $l$, $\norm{\rR\tc{l}} \leq c$.
\end{lemma}

\begin{theorem} \label{theorem:mu}
Suppose that Assumptions \ref{assumpt:model1}-\ref{assumpt:model3} hold. Then, for all $i \geq 0$, we have 
\begin{align*}
\lim_{l\to\infty} \EE{\tmu_i\tc{l}} = 0_d.
\end{align*}
\end{theorem}
\begin{IEEEproof}
Let $\tilde{u}\tc{l} = \EE{\tmu\tc{l}}$, then taking expectation on both sides of \eqref{mu_recursive}, we have $\tilde{u}\tc{l}= \rQ^{(l)} \tilde{u}\tc{l-1}$, where $\rQ\tc{l} = \EE{\bQ\tc{l}}$, and we have used the independence of $\bQ\tc{l}$ and $\tmu\tc{l-1}$. The theorem now follows from Lemma \ref{lemma:Q}. 
\end{IEEEproof}

\subsection{MSD Convergence}

Since \eqref{mu_recursive} is a linear recursive equation, its mean-square stability analysis follows from similar methods used in Section 6.5 of \cite{Sayed2013a}. We note however that our analysis is complicated by the fact that $\bQ\tc{l}$ in \eqref{mu_recursive} does not have the same form as that in \cite{Sayed2013a}, which allows decomposition of the recursion into components that then makes it possible to replace the random matrix $\bQ\tc{l}$ with $\rQ\tc{l} = \EE{\bQ\tc{l}}$. In addition, we do not have the luxury of a step size parameter that can be chosen sufficiently small so that higher order terms can be ignored. Therefore, we require stronger assumptions on the distributions of $\bH_{ij}\tc{l}$ in order to ensure MSD convergence for gSPAWN.

The MSD at iteration $l$ is defined to be $\EE{\norm{\tmu\tc{l}}^2}$. Consider a positive definite matrix $W$. Let $w=\mathrm{vec}(W)$ be the vector formed by stacking the columns of $W$ together. For any vector $a$, let $\norm{a}^2_w = \norm{a}^2_W = a^* W a$. Following the same derivation as in \cite{Sayed2013a}, it can be shown that
\begin{align}\label{ms_recursion}
\EE{\norm{\tmu\tc{l}}^2_w} = \EE{\norm{\tmu\tc{l-1}}^2_{\cF\tc{l}w}} + \cR\tc{l}w,
\end{align}  
where 
\begin{align}\label{cF}
\cF\tc{l} = \EE{\left(\bQ\tc{l}\right)^T \otimes \left(\bQ\tc{l}\right)^*},
\end{align}
and
\begin{align*}
\cR\tc{l} = \left(\mathrm{vec}\left( \rR\tc{l} \rC \left(\rR\tc{l}\right)^T \right)\right)^T,
\end{align*}
with $\rC = \mathrm{diag}(\rC_{11},\ldots,\rC_{n1},\ldots,\rC_{1n},\ldots,\rC_{nn})$. The MSD at any iteration $l$ can be found by setting $W = \rI_{dn}$. We say that gSPAWN is mean-square stable if the MSD converges as $l \to \infty$.\footnote{Similar to \cite{Sayed2013a}, the mean-square error and excess mean-square error can be defined, but we do not discuss those here.}

By iterating the recursive relationship \eqref{ms_recursion}, we have
\begin{align*}
\EE{\norm{\tmu\tc{l}}^2} = \EE{\norm{\tmu\tc{0}}^2_{\prod_{i=1}^{l}\cF\tc{i}\vect{\rI_{dn}}}} + \sum_{i=1}^{l}\cR\tc{i}\prod_{j=i+1}^{l} \cF\tc{j} \vect{\rI_{dn}}.
\end{align*}
Therefore, to show mean square stability, it suffices to show that the matrices $\cF\tc{l}$ are stable for all $l$ sufficiently large because of Lemma \ref{lemma:Q}. Without any restrictions on the distributions of $\bH_{ij}\tc{l}$, $\cF\tc{l}$ may not be stable. In the following, we present some sufficient conditions for mean-square stability, which apply in various practical applications.

\begin{proposition}\label{prop:Hconstant}
Suppose that Assumptions \ref{assumpt:model1}-\ref{assumpt:model3} hold, and $\bH_{ij}\tc{l} = \brH_{ij}$ is non-random for all $i,j$ and $l$. Then, the MSD in the gSPAWN algorithm converges.
\end{proposition}
\begin{IEEEproof}
From \eqref{cF}, the eigenvalues of $\cF\tc{l}$ are of the form $\phi\xi^*$, where $\phi$ and $\xi$ are eigenvalues of $\bQ\tc{l}$, which is assumed here to be non-random. The proposition now follows from Assumption \ref{assumpt:model3}.  
\end{IEEEproof}

\begin{proposition}
Suppose that Assumptions \ref{assumpt:model1} and \ref{assumpt:model2} hold. In addition, there exists a unitary matrix $\rU$ so that for all $i, j = 0, \ldots, n$, there exists a unitary matrix $\rV_{ij}$ such that 
\begin{enumerate}[(i)]
	\item $\rG_{ij} = \rV_{ij} \rA_{ij} \rU^*$, where $\rA_{ij}$ is a diagonal matrix with non-negative real numbers on the diagonal;
	\item $\bH_{ij}\tc{l} = \rV_{ij} \bB_{ij}\tc{l} \rU^*$ almost surely for all $l$, where $\bB_{ij}\tc{l}$ is a diagonal matrix. Furthermore, $\rB_{ij} = \EE{\bB_{ij}\tc{l}}$ and $\tilde{\rB}_{ij} = \EE{(\bB_{ij}\tc{l})^2}$ are diagonal matrices with non-negative real numbers on the diagonal, with $\tilde{\rB}_{ij} \leq \rA_{ij}^2$; and
	\item $\rC_{ij} = \rV_{ij} \rD_{ij} \rV_{ij}^*$, where $\rD_{ij}$ is a diagonal matrix with positive real numbers on the diagonal.
\end{enumerate}
Then, the MSD in the gSPAWN algorithm converges.
\end{proposition}
\begin{IEEEproof}
The same proof in Proposition \ref{prop:sufficient} shows that for all $i \geq 0$ and $l \geq 0$, $\rP_i\tc{l}$ has the form $\rU \Lambda_{i}\tc{l} \rU^*$, where $\Lambda_i\tc{l}$ is a non-negative diagonal matrix. For fixed $l\geq 0$ and $a\geq 0$, we have
\begin{align*}
\prod_{k=a}^{l} \cF\tc{k} \vect{\rI_{dn}} = \vect{\rF_l\tc{a}},
\end{align*}
where $\rF_l\tc{k}$ is defined recursively as $\rF_l\tc{k} = \EE{\left(\bQ\tc{k}\right)^* \rF_l\tc{k+1} \bQ\tc{k}}$, with $\rF_l\tc{l+1} =  \rI_{dn}$. Let $\bar{\rU} = \textrm{diag}(\rU, \ldots, \rU)$ and $\rJ\tc{k}$ be matrices consisting of $n\times n$ blocks of size $d\times d$, where the $(i,j)$-th block of $\rJ\tc{k}$ is given by
\begin{align*}
\rJ\tc{k}(i,j) =
\begin{cases}
\Lambda_{i}\tc{k}\rA_{ij}  \left(\rD_{ij} + \rB_{ij}\Lambda_j\tc{k-1}\rB_{ij} \right)^{-1} \rA_{ij}, & \text{if } j \in \stB_i, \\
0\cdot \rI_d, & \textrm{otherwise}.
\end{cases} 
\end{align*}
Viewing $\rF_l\tc{k}$ as a $n\times n$ block matrix, we show by backward induction on $k$ that each block $\rU^* \rF_l\tc{k}(i,j) \rU$ is a non-negative diagonal matrix and $\bar{\rU}^* \rF_l\tc{k} \bar{\rU} \leq (\rJ\tc{k})^* \bar{\rU}^*\rF_l\tc{k+1}\bar{\rU} \rJ\tc{k}$. It then follows from the Perron-Frobenius theorem \cite{Seneta1981} that since $\rJ\tc{k}$ has spectral radius less than one, $\rF_l\tc{a} \to 0\cdot \rI_{dn\times dn}$ exponentially fast as $l\to \infty$ so that the proposition holds.

Suppose the claim holds for $k+1$. Let $\tilde{\Pi}_{ij}\tc{k} = \left(\rD_{ij} + \rB_{ij} \Lambda_j\tc{k} \rB_{ij}\right)^{-1}$, which is a non-negative diagonal matrix for all $k$. It can be shown algebraically that 
\begin{align*}
\rU^* \rF_l\tc{k}(i,j) \rU 
& = \sum_{q=1}^n \sum_{q'=1}^n \EE{\bB_{iq}\tc{k} \tilde{\Pi}_{iq}\tc{k-1} \rA_{iq} \rU^* \rF_l\tc{k+1}(q,q') \rU  \rA_{jq'} \tilde{\Pi}_{jq'}\tc{k-1} \bB_{jq'}\tc{k}} \\
& \leq \sum_{q=1}^n \sum_{q'=1}^n\rA_{iq}\tc{k} \tilde{\Pi}_{iq}\tc{k-1} \rA_{iq} \rU^* \rF_l\tc{k+1}(q,q') \rU  \rA_{jq'} \tilde{\Pi}_{jq'}\tc{k-1} \rA_{jq'}\tc{k} \\
& = \left((\rJ\tc{k})^* \bar{\rU}^*\rF_l\tc{k+1}\bar{\rU} \rJ\tc{k}\right)(i,j),
\end{align*} 
where the inequality holds because all the matrices are non-negative and diagonal, $\EE{\bB_{iq}\tc{k}\bB_{jq'}\tc{k}} = \EE{\bB_{iq}\tc{k}}\EE{\bB_{jq'}\tc{k}} \leq \rA_{iq}\rA_{jq'}$ if $i\neq j$ or $q \neq q'$, and $\EE{(\bB_{iq}\tc{k})^2} \leq \rA_{iq}^2$ by assumption. The claim trivially holds for $k=l$, and the induction is complete. The proposition is now proved.
\end{IEEEproof}

\subsection{Numerical Examples and Discussions} \label{subsec:num}
In this section, we provide numerical results to compare the performance of gSPAWN and the ATC algorithm. For the network graph $\mathcal{G}$, we use a $k$-regular graph with $8$ nodes generated using results from \cite{Meringer1999}. Each sensor $i$ has a $2\times 1$ local parameter $s_i$ to be estimated. The matrices $\rG_{ij}$ and $\bH_{ij}\tc{l}$ are $1\times 2$ vectors. For each $i,j=0,\ldots,7$, we let $\rG_{ij}$ to be either $[15,0]$ or $[0,15]$ randomly so that Assumption \ref{assumpt:model1}\eqref{it:GG} is satisfied. The expectation $\brH_{ij}$ is chosen to be either $[10,0]$ or $[0,10]$ randomly, and each $\bH_{ij}\tc{l}$ is generated from a Gaussian distribution with covariance $2\rI_2$. The parameter $\alpha$ in gSPAWN is chosen to be $10$. The relative degree-variance in \cite{Sayed2013} is used as the combination weight in the ATC algorithm. The step-size in the ATC algorithm for sensor $i$ (cf. equation \eqref{ATCstep}) is chosen to be $\eta/r(\rR_i)$, where $0< \eta <2$.

The root mean squared error (RMSE) for each sensor $i$ is given by $\sqrt{\EE{\norm{s_i-\hat{\bs}_i\tc{l}}_2^2}}$, where $\hat{\bs}_i\tc{l}$ is the local estimate of $s_i$ in both the gSPAWN and ATC algorithm. The iteration $l$ is chosen to be sufficiently large so that the RMSE is within $10^{-5}$ of the steady state value, at which point we say that the algorithm has converged. We use the average RMSE over all sensors as the performance measure. The Cramer-Rao lower bound (CRLB) is used as a benchmark. The likelihood function $\mathcal{L}\tc{l}\left(\{\bs_i\}_{i=1}^{n}\right)$ at iteration $l$ follows from \eqref{linear} as
\begin{align}
\mathcal{L}\tc{l}\left(\{s_i\}_{i=1}^{n}\right) \propto \exp\left\{-\frac{1}{2}(\bE\tc{l} s - \bD\tc{l})^*\rC^{-1}(\bE\tc{l} s - \bD\tc{l})\right\}, \label{crlb}
\end{align}
where $s = [s_1^T, \cdots, s_{n}^T]^T$, $\bE\tc{l}$ is a selection matrix consisting of all system matrices $\{\rG_{ij},\bH_{ij}\tc{l}\}$ at iteration $l$, and $\bD\tc{l}$ and $\rC$ are the matrices obtained by stacking all measurements $\bd_{ij}\tc{l}$ and noise covariances $\rC_{ij}$ together in the order dictated by $\bE\tc{l}$. It is well known that Fisher information matrix for \eqref{crlb} is given by ${\bE\tc{l}}^*\rC^{-1}\bE\tc{l}$ \cite{Kay}. The average CRLB is obtained by averaging over $l$. 
\begin{figure}[!t]
  \centering
  \subfigure[Graph degree $k$ v.s. average RMSE]{
    \includegraphics[width=0.48\textwidth]{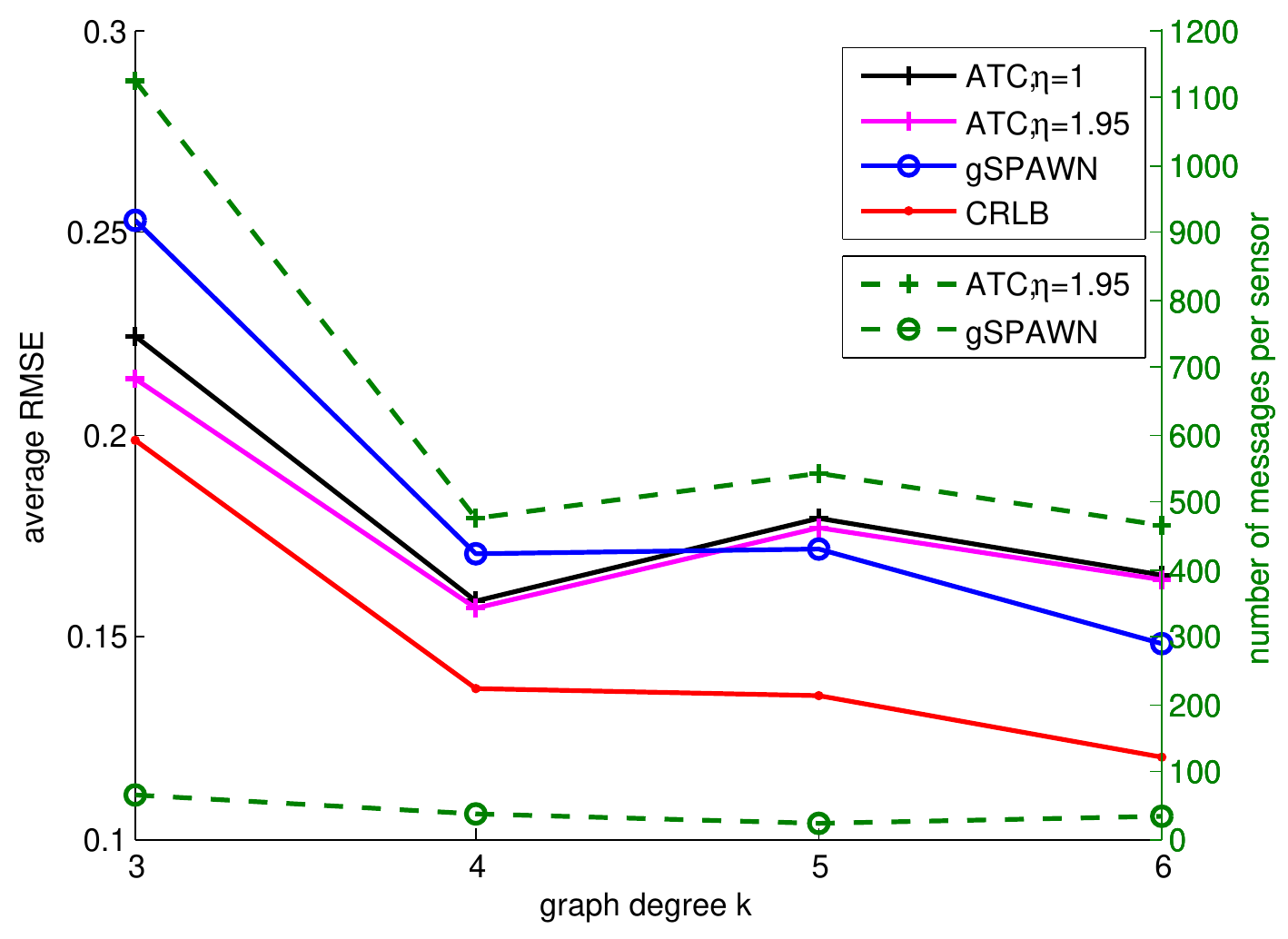}\label{Figure-RMSEReg}} 
  \subfigure[Convergence comparison.]{
    \includegraphics[width=0.48\textwidth]{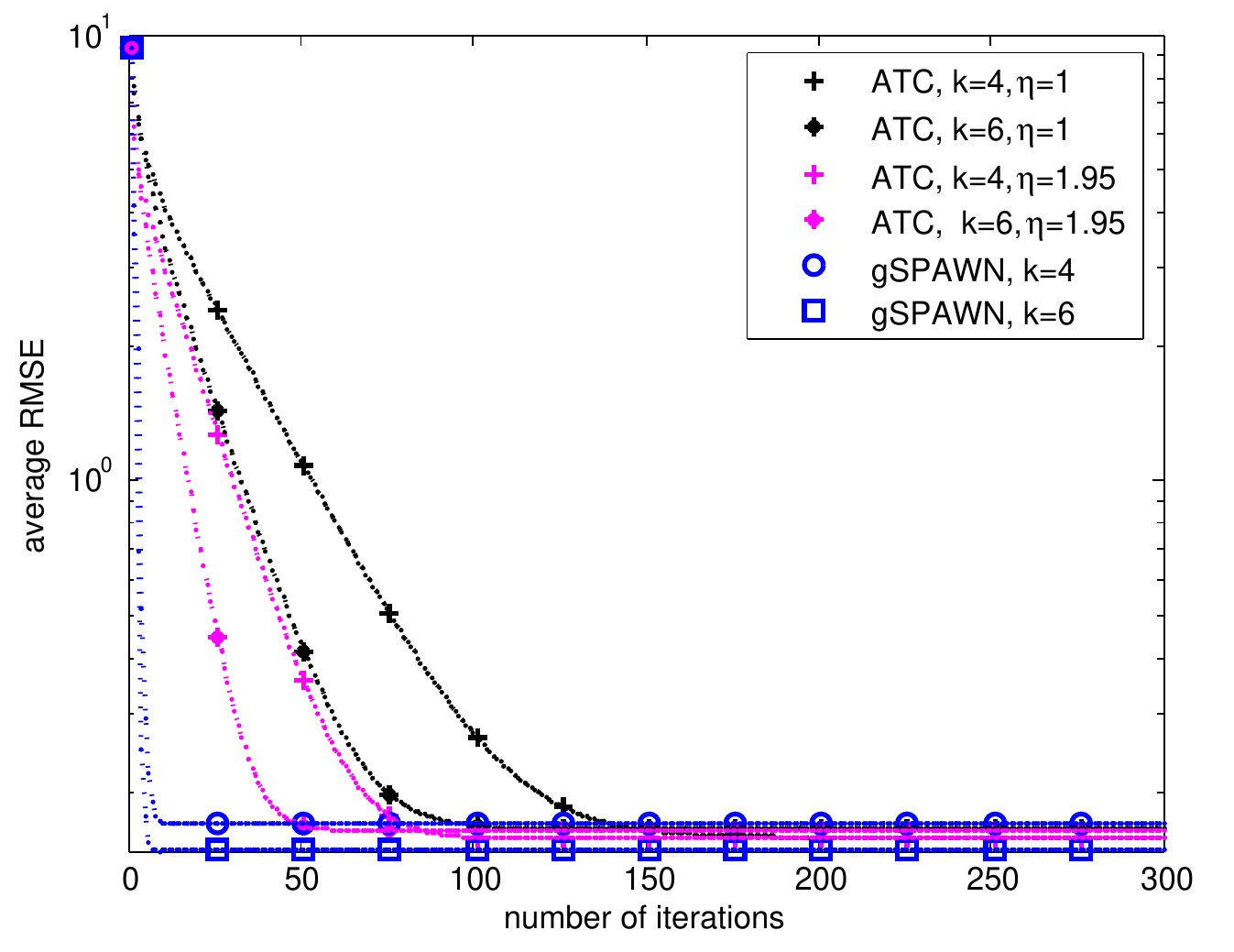}\label{Figure-convReg}} 
  \caption{Performance comparison between gSPAWN and ATC for regular graphs.} \label{Figure-k}
\end{figure}

In Figure \ref{Figure-RMSEReg}, we see that gSPAWN and ATC have comparable average RMSE when the graph degree is greater than 3. A larger graph degree results in a better average RMSE for gSPAWN, whereas the ATC algorithm performance remains relatively flat once the graph degree is greater than 3. We also observe that the total number of messages transmitted per sensor for gSPAWN is much smaller than that for ATC. This is because gSPAWN not only converges faster, but also passes fewer messages in each iteration. To be specific, the number of messages per sensor per iteration for gSPAWN is 2, and that for ATC is the graph degree $k$. 

In Figure \ref{Figure-convReg}, we observe that the gSPAWN algorithm converges within 30 iterations, while the ATC algorithm convergence rate depends not only on the graph degree but also on the value of $\eta$, with a lower graph degree and a smaller value of $\eta$ leading to a slower convergence rate. This can also be seen by comparing the average spectral radius\footnote{The spectral radius for gSPAWN is calculated for $\bQ\tc{l}$ in \eqref{Q}, and that for ATC is calculated for the recursion matrix in equation (246) of \cite{Sayed2013}.}. In our simulations, the spectral radius for ATC with $\eta=1.95$ is between 0.8905 and 0.9879, while that of the gSPAWN algorithm is between 0.5315 and 0.6711.

\section{Application to Cooperative Localization in NLOS Environments}\label{Section:Case}
\begin{figure}[!t]
\centering
\includegraphics[width=0.4\textwidth]{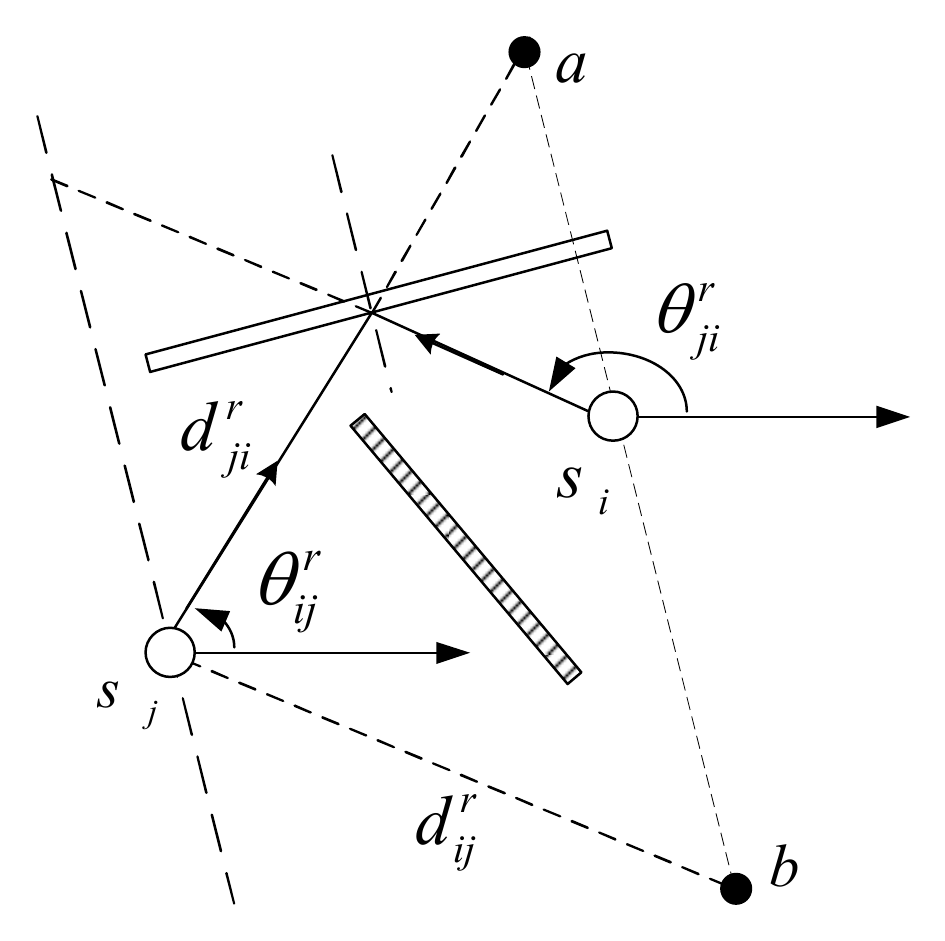}
\caption{An example for single-bounce scattering path between sensor $i$ and sensor $j$.} \label{Figure-Scatter}
\end{figure}

In this section, we apply the gSPAWN algorithm to the problem of distributed sensor localization in a NLOS environment. Suppose that there are $n+1$ sensors $0,\ldots,n$ in a network, where the position of sensor $i$ is $s_i = [x_i, y_i]^T$ with $x_i$ and $y_i$ being the $x$ and $y$ coordinates respectively. Without loss of generality, we assume that the position of the reference sensor $0$ is known up to some estimation accuracy. The objective of each sensor $i$ is to perform self-localization relative to sensor $0$ by measuring time-of-arrival (TOA) or received signal strength, and angle-of-arrival (AOA) of signals broadcast by each sensor. In most urban and indoor environments, sensors may not have direct LOS to each other. Signals transmitted by one sensor may be propagated to another sensor through multiple paths. The multipath propagation between sensors can be characterized by a ray tracing model \cite{Seow2008}. In this application, we restrict ourselves to paths that have at most one reflection. An example of a single-bounce scattering path is shown in Figure \ref{Figure-Scatter}, where a signal transmitted from sensor $j$ to sensor $i$ is reflected at a nearby scatterer. Suppose that there are $k_{ij}$ LOS or NLOS paths between sensor $i$ and sensor $j$, and that individual paths can be resolved from the received signal at each node. This is the case if, for example, the signal bandwidth is sufficiently large.

Consider the $k$th path between sensor $i$ and $j$ as shown in Figure \ref{Figure-Scatter}. Let $\bd_{ij}^{k}$ be the distance measured by sensor $i$ using the TOA or received signal strength along this path from sensor $j$, and $\theta_{ij}^k$ be the corresponding AOA of the signal. We assume that a particular direction has been fixed to be the horizontal direction, and all angles are measured w.r.t.\ this direction. In addition, sensors do not have prior knowledge of the positions of the scatterers. It can be shown using geometrical arguments \cite{Seow2008, Leng2012} that
\begin{align}\label{relation}
\bd_{ij}^k & =
g(\theta_{ij}^k, \theta_{ij}^k)^{T} \left(s_i - s_j\right),
\end{align}
where
\begin{align}\label{eq:g}
g(\theta, \phi) & =
\begin{bmatrix}
\frac{\sin\theta + \sin\phi}{\sin(\theta - \phi)} \\
-\frac{\cos\theta + \cos\phi}{\sin(\theta - \phi)}
\end{bmatrix}.
\end{align}
When the signal path between sensor $i$ and $j$ is a LOS path, we have $\vert \theta_{ij}^k - \theta_{ij}^k \vert  = \pi$ and $d_{ij}^k = \Vert s_i - s_j \Vert$, therefore we define $g\left(\theta_{ij}^k, \theta_{ij}^k\right) = [\cos(\theta_{ij}^k), \sin(\theta_{ij}^k)]^T$ in \eqref{relation}.

We stack the measurements from all $k_{ij}$ paths into vectors $\tilde{\bd}_{ij} = [\bd_{ij}^1, \ldots, \bd_{ij}^{k_{ij}}]^T$ and $\rT_{ij} = [g(\theta_{ij}^1, \theta_{ij}^1), \ldots, g(\theta_{ij}^{k_{ij}}, \theta_{ij}^{k_{ij}})]^T$. We assume that $k_{ij}>1$ and $\rT_{ij}$ has linearly independent columns, otherwise some of the paths are redundant as they do not provide additional information to perform localization. Let $\bd_{ij} = \rT_{ij}^{\dagger}\tilde{\bd}_{ij}$, where $\rT_{ij}^\dagger$ is the Moore-Penrose pseudo-inverse of $\rT_{ij}$. We adopt the data model
\begin{align}
\bd_{ij} = s_i - s_j + \bomega_{ij}, \label{localization}
\end{align}
where we assume that the AOA of each arriving signal can be measured sufficiently accurately \cite{Andersen2002} so that the distance and AOA measurement noise can be included together in $\bomega_{ij}$, which is a $2 \times 1$ measurement noise vector with covariance $\rC_{ij} = \sigma_{ij}^2(\rT_{ij}^{\dagger})(\rT_{ij}^{\dagger})^T = \sigma_{ij}^2(\rT_{ij}^T\rT_{ij})^{-1}$. Note that $\rC_{ij}$ is a function of the AOA of the signal from sensor $j$ to sensor $i$.

This corresponds to the general model in \eqref{linear}, where sensor observations are the same over all iterations $l$, and $\rG_{ij}=\bH_{ij}\tc{l} = \rI_2$ for all $l$. The gSPAWN algorithm in Section \ref{Section:Algo} can now be used to estimate $s_i$ in a distributed manner. In this non-adaptive application, we assume that only one measurement is observed at each sensor, but the same algorithm can also be applied if new observations are available at each iteration of the gSPAWN algorithm as shown in Section \ref{Section:Algo}. The local updates \eqref{P} and \eqref{mu} at sensor $i$ in the $l$th iteration are given by
\begin{align}
\rP_i\tc{l} & = \left(\sum_{j\in\stB_i} \left(\rC_{ij} + \rP_j\tc{l-1}\right)^{-1}\right)^{-1}, \label{locP}\\
\bmu_i\tc{l} & = \rP_i\tc{l} \sum_{j\in\stB_i} \left(\rC_{ij} + \rP_j\tc{l-1}\right)^{-1}\left(\bmu_j\tc{l-1}+\bd_{ij}\right), \label{locmu}
\end{align}
where $\bmu_i\tc{l}$ is the sensor's local estimate of its position after $l$ iterations.

\subsection{Convergence of Location Estimates}
Suppose that all scatterers in the environment are either parallel or orthogonal to each other (e.g., in indoor environments, the strongest scatterers are the walls, floor and ceiling). The convergence of the gSPAWN algorithm in distributed localization of the sensors then follows from results in Section \ref{Section:convergence}. 

\begin{corollary}\label{case}
In the gSPAWN algorithm for the NLOS localization problem in a strongly connected sensor network using data model \eqref{localization}, we have
\begin{enumerate}[(i)]
\item \label{case-P} the covariance matrix in \eqref{locP} converges;
\item \label{case-U} the belief means in \eqref{locmu} converges in mean to the true sensor locations when all scatterers are either parallel or orthogonal to each other, and
\item \label{case-M} the location MSDs converge when all scatterers are either parallel or orthogonal to each other.
\end{enumerate}
\end{corollary}
\begin{IEEEproof}
From \eqref{localization}, it can be checked that Assumptions \ref{assumpt:model1} and \ref{assumpt:model2} are satisfied with $\rG_{ij} = \brH_{ij} = \bH_{ij}\tc{l} = \rI_2$. Claim \eqref{case-P} now follows from Theorem \ref{theorem:P}. We next show that Assumption \ref{assumpt:model3} holds when all scatterers are either parallel or orthogonal to each other. In this case, there exists a unitary matrix $\rU$ such that after applying $\rU$ to the system model, all the scatterers are either horizontal or vertical. Then for all $i=1,\ldots,n$, $j \in \stB_i$, and each path $k$, we have $(\theta_{ij}^{k} + \theta_{ij}^{k} - 2\rho) \ (\textrm{mod} \ 2\pi)  = \pi$, where $\rho = 0$ or $\pi/2$ depending on whether the path is reflected by a horizontal or vertical scatterer, respectively. From \eqref{eq:g}, we obtain
\begin{align*}
\rU^*\rC_{ij}\rU = \sigma_{ij}^2
\begin{bmatrix}
\sum_{k \in \mathcal{K}_h} [\sec(\theta_{ij}^k)]^2 & 0 \\
0 & \sum_{k \in \mathcal{K}_v} [\csc(\theta_{ij}^k)]^2
\end{bmatrix}^{-1},
\end{align*}
where $\mathcal{K}_h$ and $\mathcal{K}_v$ are the index sets of paths from horizontal and vertical scatterers, respectively. Assumption \ref{assumpt:model3} now follows from Proposition \ref{prop:sufficient}, and claims \eqref{case-U} and \eqref{case-M} then follow from Theorem \ref{theorem:mu} and Proposition \ref{prop:Hconstant} respectively. The corollary is now proved.
\end{IEEEproof}

\subsection{Simulation Results and Discussions}\label{subsect:Sim}
\begin{figure}[!t]
  \centering
  \includegraphics[width=0.48\textwidth]{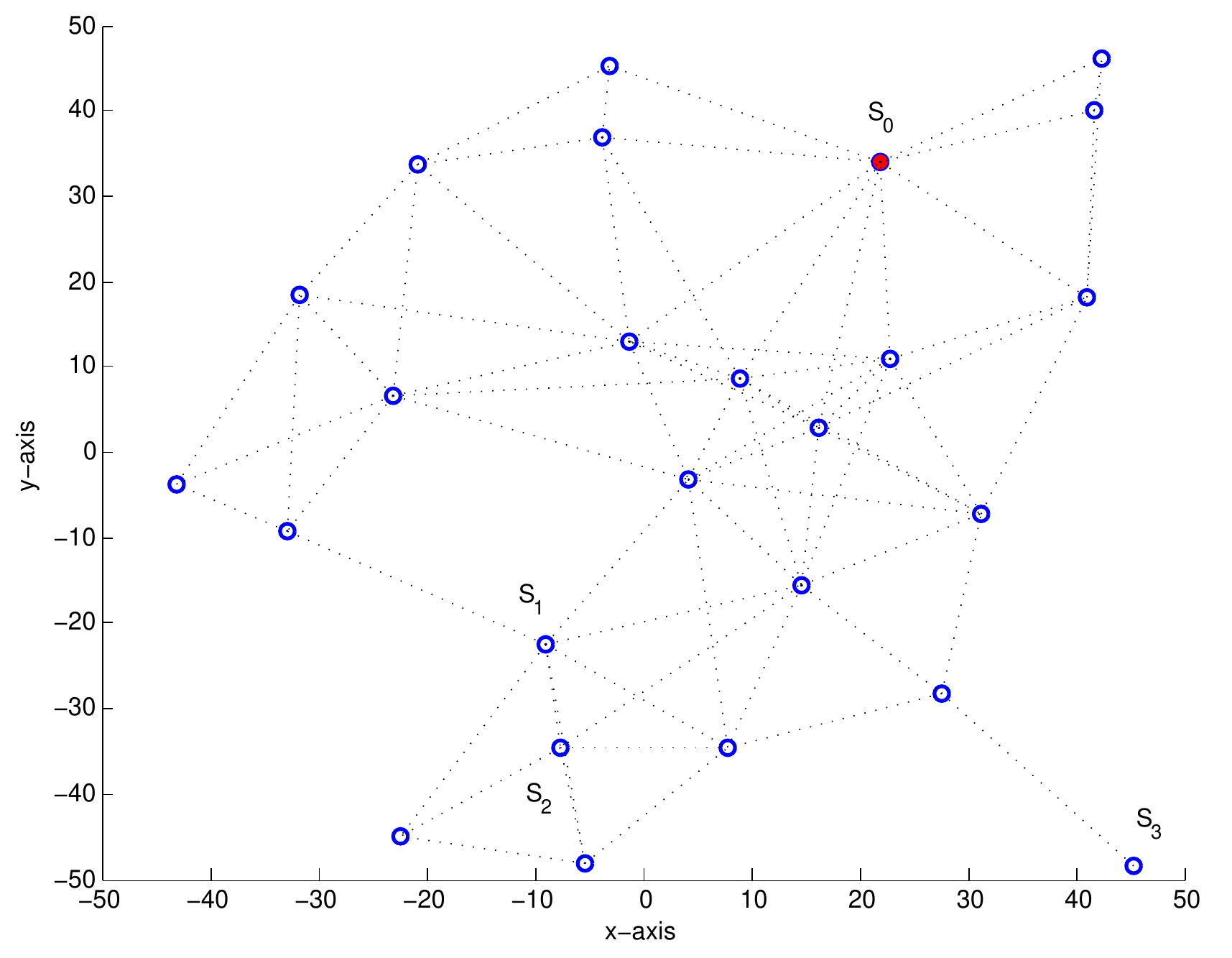}
  \caption{A signal graph corresponding to a network of 25 nodes. The reference node is $s_0$.}
  \label{Figure-Node}
\end{figure}

In this section, we present simulation results for the distributed localization problem. We consider a network with $25$ nodes scattered in an enclosed area of $100 \mathrm{m} \times 100 \mathrm{m}$. The communication links between nodes are shown in Figure \ref{Figure-Node}. Scatterers are placed throughout the area. Simulations are conducted with different values of the measurement noise parameter $\sigma_{ij} = \sigma$ for all $i,j$. Each simulation run consists of 1000 independent trials. We use the average RMSE over all sensors as the performance measure, and we also use CRLB as a benchmark, which are calculated in a way similar to that in Section \ref{subsec:num}. 

We compare the performance of gSPAWN with that of a peer-to-peer localization method and the ATC method. In the peer-to-peer localization approach, sensors that have direct measurements w.r.t.\ the reference node are localized first, and then other sensors are localized by treating previously localized neighbors as virtual anchor nodes. In the ATC strategy, the step-size parameter for $s_i$ is set to be $1/r(\rR_i)$, where $r(\rR_i)$ is as defined in \eqref{ATCstep}. The relative degree-variance is used as combination weights. 

\begin{figure}[!t]
  \centering
  \subfigure[noise parameter v.s. average RMSE]{
    \includegraphics[width=0.48\textwidth]{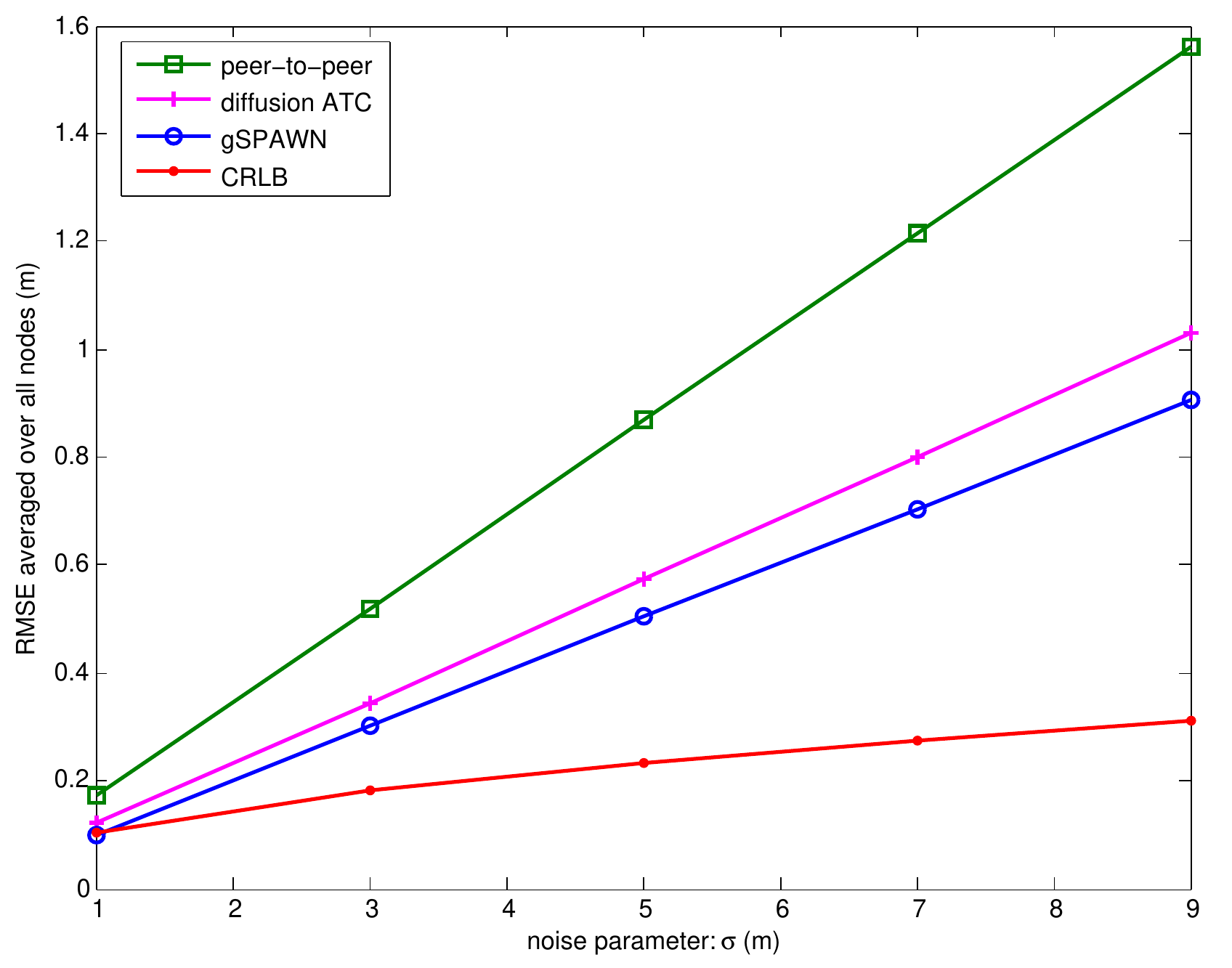}\label{Figure-RMSE}} 
  \subfigure[Convergence performance with $\sigma=1$]{
    \includegraphics[width=0.48\textwidth]{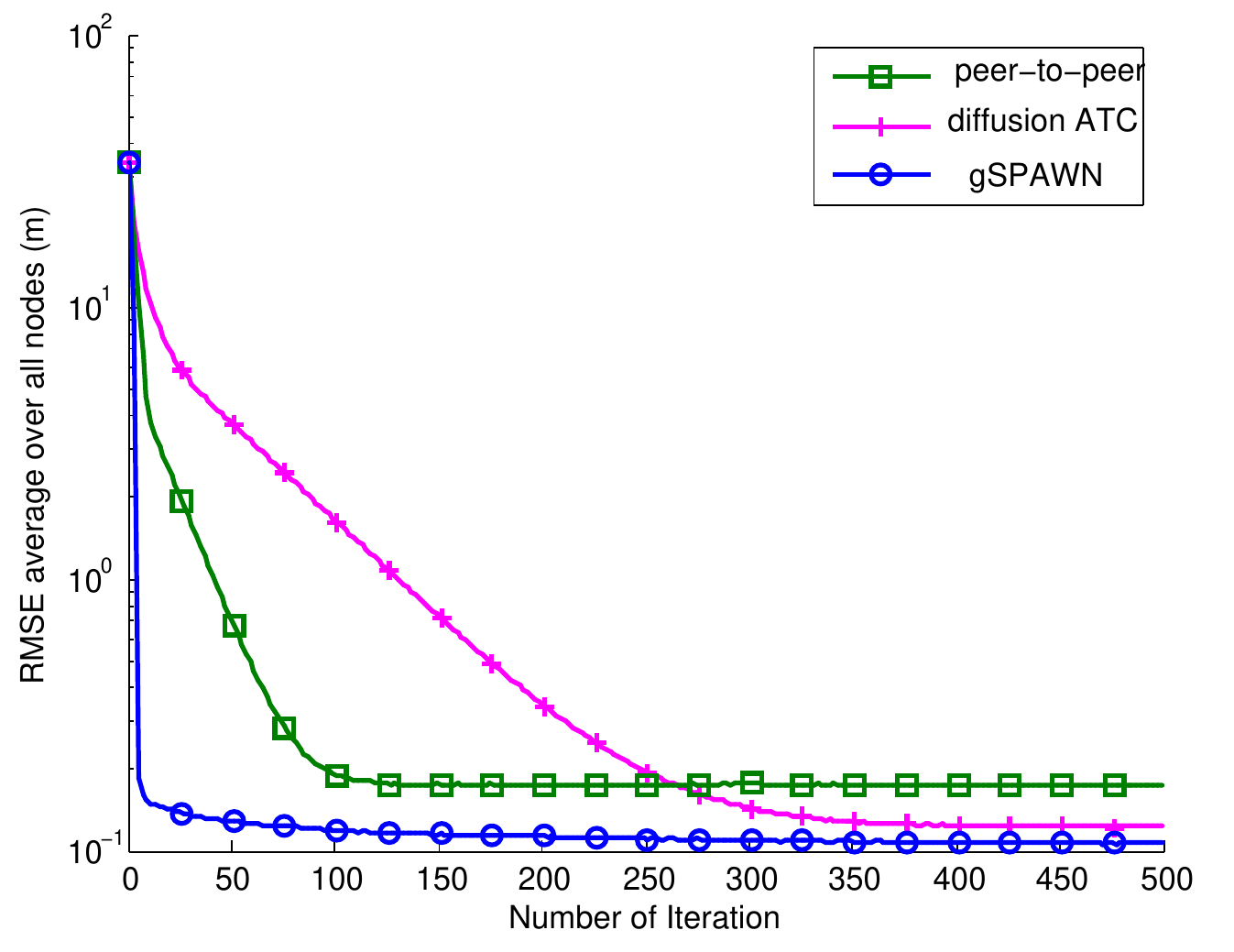}\label{Figure-Conv}} 
  \caption{Performance comparison betweenn gSPAWN, diffusion ATC, and peer-to-peer algorithm.}
\end{figure}

In Figure  \ref{Figure-RMSE}, we show the average RMSE with respect to different values of $\sigma$. It can be seen that the gSPAWN algorithm outperforms both ATC and the peer-to-peer localization method, with average RMSE close to that of the ATC strategy. The convergence performance is shown in Figure \ref{Figure-Conv}, where $\sigma=1$. The gSPAWN algorithm converges within 100 iterations, while the ATC converges only after 350 iterations. In this numerical example, the peer-to-peer method uses one broadcast per sensor per iteration, the gSPAWN algorithm uses two broadcasts, i.e., one for the estimated mean and the other for the covariance information, per sensor per iteration, while the ATC strategy uses an average of 5.68 broadcasts per sensor per iteration.

\section{Conclusions}\label{Section:Conclude}
We have studied a distributed local linear parameter estimation algorithm called gSPAWN, based on the sum-product algorithm. The gSPAWN algorithm lets each node in the network broadcast a fixed size message to all neighbors in each iteration, with the message size remaining constant regardless of the network size and density. It is therefore well suited for implementation in WSNs. We show that the gSPAWN algorithm converges in mean, and has mean-square stability under some technical sufficient conditions. We also show how to apply the gSPAWN algorithm to a network localization problem in NLOS environments. Our simulation results suggest that gSPAWN has better average RMSE performance than the diffusion ATC and peer-to-peer localizaton methods. As part of future research work, it would be of interest to investigate the convergence properties of the more general SPAWN algorithm, which can be used for non-linear local parameter estimation, and which also has the advantage of broadcasting a fixed size message per iteration per sensor.

\appendices
\section{Numerical Example of $r(\rQ\tc{\infty})>1$.}\label{appendix:counterexample}
In this appendix, we provide a numerical example that shows that unlike the case where measurements are scalar with $m=d=1$, the matrix $\rQ\tc{\infty}$ given by \eqref{Qinfty} may have spectral radius greater than one if additional conditions are not imposed. We consider a linear graph with $n=2$ nodes and a reference node $0$. There is an edge between node $0$ and node $1$, and another edge between node $1$ and node $2$. Let $\rG_{ij} = \brH_{ij} = \rI_2$ for all $i,j$ in \eqref{Qinfty}, $\alpha_0 = 0$, $\rC_{01} = 100 \cdot \rI_2$, and for $i,j=1,2$, $\rC_{ij} = \rA_{ij}^{-1} - \rP_j\tc{\infty}$, where
\begin{align*}
 & \rA_{11}  =
\begin{bmatrix}
    0.3586 &  -0.1580 \\
   -0.1580 &   0.1135 
\end{bmatrix},
 \rA_{12}  =
\begin{bmatrix}
    0.0709 &  -0.0890 \\
   -0.0890 &   0.1118
\end{bmatrix}, \\
& \rA_{21} =
\begin{bmatrix}
    0.3670 &  -0.2359 \\
   -0.2359 &   0.1522
\end{bmatrix},
 \rA_{22} =
\begin{bmatrix}
    0.3067 &  -0.0163 \\
   -0.0163 &   0.0357
\end{bmatrix},
\end{align*}
and 
\begin{align*}
 & \left(\rP_{1}\tc{\infty}\right)^{-1}  =
\begin{bmatrix}
    0.4395 &  -0.2470 \\
   -0.2470 &   0.2353
\end{bmatrix},
\left(\rP_{2}\tc{\infty}\right)^{-1}  =
\begin{bmatrix}
    0.6737 &  -0.2522 \\
   -0.2522 &   0.1879
\end{bmatrix}.
\end{align*}
It can be checked that $\rP_{1}\tc{\infty}$ and $\rP_{2}\tc{\infty}$ satisfy \eqref{Pinfty}, and
\begin{align*}
 & \rQ\tc{\infty}  =
\begin{bmatrix}
    1.0695  & -0.2156 &  -0.1250 &  0.1574\\
    0.4512  &  0.2560 &  -0.5094 &  0.6403\\
    0.1503  & -0.0943 &   0.8497 &  0.0943\\
   -1.0537  &  0.6834 &   1.0537 &  0.3166
\end{bmatrix},
\end{align*}
which has a spectral radius of $1.017$.

\bibliographystyle{IEEEtran}
\bibliography{IEEEabrv,Distributed_Localization}

\begin{thebibliography}{10}
\providecommand{\url}[1]{#1}
\csname url@samestyle\endcsname
\providecommand{\newblock}{\relax}
\providecommand{\bibinfo}[2]{#2}
\providecommand{\BIBentrySTDinterwordspacing}{\spaceskip=0pt\relax}
\providecommand{\BIBentryALTinterwordstretchfactor}{4}
\providecommand{\BIBentryALTinterwordspacing}{\spaceskip=\fontdimen2\font plus
\BIBentryALTinterwordstretchfactor\fontdimen3\font minus
  \fontdimen4\font\relax}
\providecommand{\BIBforeignlanguage}[2]{{%
\expandafter\ifx\csname l@#1\endcsname\relax
\typeout{** WARNING: IEEEtran.bst: No hyphenation pattern has been}%
\typeout{** loaded for the language `#1'. Using the pattern for}%
\typeout{** the default language instead.}%
\else
\language=\csname l@#1\endcsname
\fi
#2}}
\providecommand{\BIBdecl}{\relax}
\BIBdecl

\bibitem{Akyildiz2007}
I.~F. Akyildiz, T.~Melodia, and K.~R. Chowdury, ``Wireless multimedia sensor
  networks: A survey,'' \emph{{IEEE} Wireless Commun. Mag.}, vol.~14, no.~6,
  pp. 32--39, 2007.

\bibitem{Bulusu2005}
N.~Bulusu and S.~Jha, \emph{Wireless Sensor Networks: A Systems
  Perspective}.\hskip 1em plus 0.5em minus 0.4em\relax Artech House, 2005.

\bibitem{Tay2009}
W.~P. Tay, J.~N. Tsitsiklis, and M.~Z. Win, ``Bayesian detection in bounded
  height tree networks,'' \emph{{IEEE} Trans. Signal Process.}, vol.~57,
  no.~10, pp. 4042--4051, Oct. 2009.

\bibitem{Tay2008}
------, ``On the impact of node failures and unreliable communications in dense
  sensor networks,'' \emph{{IEEE} Trans. Signal Process.}, vol.~56, no.~6, pp.
  2535--2546, Jun. 2008.

\bibitem{Hong2007}
Y.-W. Hong, W.-J. Huang, and F.-H. Chiu, ``Cooperative communications in
  resource-constrained wireless networks,'' \emph{{IEEE} Signal Process. Mag.},
  vol.~24, no.~3, pp. 47--57, 2007.

\bibitem{Zhu2010}
H.~Zhu, A.~Cano, and G.~B. Giannakis, ``Distributed consensus-based
  demodulation: algorithms and error analysis,'' \emph{{IEEE} Trans. Wireless
  Commun.}, vol.~9, no.~6, pp. 2044--2054, 2010.

\bibitem{Speranzon2006}
A.~Speranzon, C.~Fischione, and K.~H. Johansson, ``Distributed and
  collaborative estimation over wireless sensor networks,'' in \emph{Decision
  and Control, 2006 45th IEEE Conference on}, 2006, pp. 1025--1030.

\bibitem{Gholami2012}
M.~R. Gholami, S.~Gezici, and E.~G. Strom, ``Improved position estimation using
  hybrid tw-toa and tdoa in cooperative networks,'' \emph{{IEEE} Trans. Signal
  Process.}, vol.~60, no.~7, pp. 3770--3785, 2012.

\bibitem{Dardari2008}
D.~Dardari, A.~Conti, J.~Lien, and M.~Z. Win, ``The effect of cooperation on
  uwb-based positioning systems using experimental data,'' \emph{EURASIP J.
  Adv. Signal Process}, vol. 2008, pp. 124:1--124:11, Jan. 2008.

\bibitem{Wymeersch2008}
H.~Wymeersch, U.~Ferner, and M.~Win, ``Cooperative bayesian self-tracking for
  wireless networks,'' \emph{Communications Letters, IEEE}, vol.~12, no.~7, pp.
  505--507, 2008.

\bibitem{Wymeersch2009}
H.~Wymeersch, J.~Lien, and M.~Z. Win, ``Cooperative localization in wireless
  networks,'' \emph{Proc. {IEEE}}, vol.~97, no.~2, pp. 427--450, 2009.

\bibitem{Boyd2006}
S.~Boyd, A.~Ghosh, B.~Prabhakar, and D.~Shah, ``Randomized gossip algorithms,''
  \emph{Information Theory, IEEE Transactions on}, vol.~52, no.~6, pp.
  2508--2530, 2006.

\bibitem{Aysal2009}
T.~C. Aysal, M.~E. Yildiz, A.~D. Sarwate, and A.~Scaglione, ``Broadcast gossip
  algorithms for consensus,'' \emph{{IEEE} Trans. Signal Process.}, vol.~57,
  no.~7, pp. 2748--2761, 2009.

\bibitem{Zhu2011}
S.~Zhu and Z.~Ding, ``Distributed cooperative localization of wireless sensor
  networks with convex hull constraint,'' \emph{{IEEE} Trans. Wireless
  Commun.}, vol.~10, no.~7, pp. 2150--2161, 2011.

\bibitem{Kriegleder2013}
M.~Kriegleder, R.~Oung, and R.~D'Andrea, ``Asynchronous implementation of a
  distributed average consensus algorithm,'' in \emph{IEEE/RSJ International
  Conference of Intelligent Robots and Systems}, Tokyo, Japan, Nov. 3-7 2013.

\bibitem{Lopes2007}
C.~G. Lopes and A.~H. Sayed, ``Incremental adaptive strategies over distributed
  networks,'' \emph{IEEE Trans. Signal Processing}, vol.~55, no.~8, pp.
  4064--4077, August 2007.

\bibitem{Cattivelli2011}
F.~Cattivelli and A.~H. Sayed, ``Analysis of spatial and incremental {LMS}
  processing for distributed estimation,'' \emph{IEEE Trans. on Signal
  Process.}, vol.~59, no.~4, pp. 1465--1480, April 2011.

\bibitem{Schizas2009}
I.~Schizas, G.~Mateos, and G.~Giannakis, ``Distributed lms for consensus-based
  in-network adaptive processing,'' \emph{Signal Processing, IEEE Transactions
  on}, vol.~57, no.~6, pp. 2365--2382, 2009.

\bibitem{Sayed2013a}
\BIBentryALTinterwordspacing
A.~H. Sayed, ``Diffusion adaptation over networks,'' in \emph{E-Reference
  Signal Processing}, R.~Chellapa and S.~Theodoridis, Eds.\hskip 1em plus 0.5em
  minus 0.4em\relax Elsevier, 2013. [Online]. Available:
  \url{http://arxiv.org/abs/1205.4220}
\BIBentrySTDinterwordspacing

\bibitem{Tu2012}
S.-Y. Tu and A.~Sayed, ``Diffusion strategies outperform consensus strategies
  for distributed estimation over adaptive networks,'' \emph{Signal Processing,
  IEEE Transactions on}, vol.~60, no.~12, pp. 6217--6234, 2012.

\bibitem{Sayed2013}
A.~H. Sayed, S.-Y. Tu, J.~Chen, X.~Zhao, and Z.~J. Towfic, ``Diffusion
  strategies for adaptation and learning over networks: an examination of
  distributed strategies and network behavior,'' \emph{{IEEE} Signal Process.
  Mag.}, vol.~30, no.~3, pp. 155--171, 2013.

\bibitem{Leng2011}
M.~Leng and Y.-C. Wu, ``Distributed clock synchronization for wireless sensor
  networks using belief propagation,'' \emph{{IEEE} Trans. Signal Process.},
  vol.~59, no.~11, pp. 5404--5414, 2011.

\bibitem{Kschischang2001}
F.~R. Kschischang, B.~J. Frey, and H.-A. Loeliger, ``Factor graphs and the
  sum-product algorithm,'' \emph{{IEEE} Trans. Inf. Theory}, vol.~47, no.~2,
  pp. 498--519, 2001.

\bibitem{Bishop2006}
C.~M. Bishop, \emph{Pattern Recognition and Machine Learning}.\hskip 1em plus
  0.5em minus 0.4em\relax Springer, 2006.

\bibitem{Pearl1988}
J.~Pearl, \emph{Probabilistic Reasoning in Intelligent Systems: Networks of
  Plausible Inference}, 2nd~ed.\hskip 1em plus 0.5em minus 0.4em\relax San
  Francisco, CA: Morgan Kaufmann., 1988.

\bibitem{Johnson2006}
J.~K. Johnson, D.~M. Malioutov, and A.~S. Willsky, ``Walk-sum interpretation
  and analysis of {Gaussian} belief propagation,'' \emph{Advances in Neural
  Information Processing System}, vol.~18, 2006.

\bibitem{Weiss2001}
\BIBentryALTinterwordspacing
Y.~Weiss and W.~T. Freeman, ``Correctness of belief propagation in gaussian
  graphical models of arbitrary topology,'' \emph{Neural Comput.}, vol.~13,
  no.~10, pp. 2173--2200, Oct. 2001. [Online]. Available:
  \url{http://dx.doi.org/10.1162/089976601750541769}
\BIBentrySTDinterwordspacing

\bibitem{Cattivelli2010}
F.~S. Cattivelli and A.~H. Sayed, ``Diffusion {LMS} strategies for distributed
  estimation,'' \emph{{IEEE} Trans. Signal Process.}, vol.~58, no.~3, pp.
  1035--1048, 2010.

\bibitem{Takahashi2010}
N.~Takahashi, I.~Yamada, and A.~H. Sayed, ``Diffusion least-mean squares with
  adaptive combiners: Formulation and performance analysis,'' \emph{{IEEE}
  Trans. Signal Process.}, vol.~58, no.~9, pp. 4795--4810, 2010.

\bibitem{Srirangarajan2008}
S.~Srirangarajan, A.~Tewfik, and Z.-Q. Luo, ``Distributed sensor network
  localization using {SOCP} relaxation,'' \emph{{IEEE} Trans. Wireless
  Commun.}, vol.~7, no.~12, pp. 4886--4895, 2008.

\bibitem{Ihler2005}
A.~T. Ihler, I.~Fisher, J.~W., R.~L. Moses, and A.~S. Willsky, ``Nonparametric
  belief propagation for self-localization of sensor networks,'' \emph{{IEEE}
  J. Sel. Areas Commun.}, vol.~23, no.~4, pp. 809--819, 2005.

\bibitem{Miao2007}
H.~Miao, K.~Yu, and M.~J. Juntti, ``Positioning for {NLOS} propagation:
  Algorithm derivations and {Cramer-Rao} bounds,'' \emph{{IEEE} Trans. Veh.
  Technol.}, vol.~56, no.~5, pp. 2568--2580, 2007.

\bibitem{Seow2008}
C.~K. Seow and S.~Y. Tan, ``Non-line-of-sight localization in multipath
  environments,'' \emph{{IEEE} Trans. Mobile Comput.}, vol.~7, no.~5, pp.
  647--660, 2008.

\bibitem{Xie2009}
Y.~Xie, Y.~Wang, P.~Zhu, and X.~You, ``Grid-search-based hybrid {TOA}/{AOA}
  location techniques for {NLOS} environments,'' \emph{{IEEE} Commun. Lett.},
  vol.~13, no.~4, pp. 254--256, 2009.

\bibitem{Sarhan2010}
A.~M. Sarhan, N.~M. EI-Shazly, and E.~M. Shehata, ``On the existence of
  extremal positive definite solutions of the nonlinear matrix equation
  $\mathbf{X}^r + \sum_{i=1}^{m} \mathbf{A}_i^* \mathbf{X}^{\delta_i}
  \mathbf{A}_i = \mathbf{I}$,'' \emph{Mathematical and Computer Modelling},
  vol.~51, no. 9-10, 2010.

\bibitem{Milliken1977}
G.~A. Milliken, ``A theorem on the difference of the generalized inverse of two
  nonnegative matrices,'' \emph{Commun. Statist.-Theor. Meth.}, vol. A6(1),
  1977.

\bibitem{Seneta1981}
E.~Seneta, \emph{Nonnegative matrices and markov chains}, 2nd~ed., ser.
  Springer Series in Statistics.\hskip 1em plus 0.5em minus 0.4em\relax
  Springer, 1981.

\bibitem{Meringer1999}
M.~Meringer, ``Fast generation of regular graphs and construction of cages,''
  \emph{Journal of Graph Theory}, vol.~30, pp. 137--146, 1999.

\bibitem{Kay}
S.~M. Kay, \emph{Fundamentals of Statistical Signal Processing: Estimation
  Theory}.\hskip 1em plus 0.5em minus 0.4em\relax Englewood Cliffs, NJ, 1993.

\bibitem{Leng2012}
M.~Leng, W.-P. Tay, and T.~Quek, ``Cooperative and distributed localization for
  wireless sensor networks in multipath environments,'' in \emph{Acoustics,
  Speech and Signal Processing (ICASSP), 2012 IEEE International Conference
  on}, 2012, pp. 3125--3128.

\bibitem{Andersen2002}
J.~B. Andersen and K.~I. Pedersen, ``Angle-of-arrival statistics for low
  resolution antennas,'' \emph{{IEEE} Trans. Antennas Propag.}, vol.~50, no.~3,
  pp. 391--395, 2002.

\end{thebibliography}

\end{document}